\documentclass[pdflatex,sn-mathphys-num]{sn-jnl}
\usepackage[utf8]{inputenc}
\usepackage[T1]{fontenc}
\usepackage{graphicx}%
\usepackage{multirow}%
\usepackage{amsmath,amssymb,amsfonts}%
\usepackage{amsthm}%
\usepackage{mathrsfs}%
\usepackage[title]{appendix}%
\usepackage{xcolor}%
\usepackage{textcomp}%
\usepackage{manyfoot}%
\usepackage{booktabs}%
\usepackage{algorithm}%
\usepackage{algorithmicx}%
\usepackage{algpseudocode}%
\usepackage{listings}%
\usepackage{makecell}
\usepackage{latexsym}
\usepackage{array}
\usepackage{tabularx}
\usepackage{multirow}
\usepackage{amsfonts}
\usepackage{pifont}
\usepackage{array}
\usepackage{tikz}
\usetikzlibrary{tikzmark,calc}
\usepackage{colortbl}
\usepackage{xcolor}
\usepackage{pst-all} 
\usepackage{pst-poly}
\newpsobject{showgrid}{psgrid}{subgriddiv=1,griddots=10,gridlabels=6pt}
\usepackage{graphicx}
\usepackage{fancybox}
\usepackage{wrapfig}
\usepackage{todonotes}
\graphicspath{{./}{./fig/}}

\usepackage{caption}
\newcommand{\con}{\wedge} 
\newcommand{\dis}{\vee} 
\newcommand{\alw}{\Box} 
\newcommand{\imp}{\Rightarrow} 
\newcommand{\som}{\Diamond} 

\theoremstyle{thmstyleone}%
\newtheorem{theorem}{Theorem}
\theoremstyle{thmstyletwo}%
\newtheorem{fact}[theorem]{Fact} 
\newtheorem{remark}{Remark}%

\theoremstyle{thmstylethree}%
\newtheorem{definition}{Definition}%

\begin{document}

\title[Logic Mining from Process Logs]
      {Logic Mining from Process Logs: Towards Automated Specification and Verification}


\author{\centering Rados{\l}aw Klimek \& Julia Witek\\
AGH University of Krakow\\
rklimek@agh.edu.pl}



\abstract{Logical specifications play a key role in the formal analysis of behavioural models.
Automating the derivation of such specifications is particularly valuable in complex systems,
where manual construction is time-consuming and error-prone.
This article presents an approach for generating logical specifications from
process models discovered via workflow mining,
combining pattern-based translation with automated reasoning techniques.
In contrast to earlier work,
we evaluate the method on both general-purpose and real-case event logs,
enabling a broader empirical assessment.
The study examines the impact of data quality,
particularly noise, on the structure and testability of generated specifications.
Using automated theorem provers, we validate a variety of logical properties,
including satisfiability, internal consistency, and alignment with predefined requirements.
The results support the applicability of the approach in realistic settings and
its potential integration into empirical software engineering practices.}

\maketitle

\section{Introduction}


Formal logic has long been recognised as a foundation of computer science, providing rigorous tools for modelling, analysing, and verifying system behaviour~\cite{DeMol-Primiero-2015}. Logical specifications, in particular, play a central role in ensuring clarity and correctness throughout the software development lifecycle~\cite{Broy-2013}. They support not only formal verification and automated reasoning, but also enhance communication within teams by offering unambiguous representations of system dynamics.
In practice, however, the effort required to manually construct such specifications,
especially for large or complex models,
often limits their use.

The present work investigates the automation of logical specification generation, building upon the method introduced in~\cite{Klimek-2019-LAMP}, which addressed this problem for activity diagrams through a pattern-based translation framework. We extend this line of research by shifting focus from design-time models to observational data, specifically process logs extracted via workflow mining~\cite{vanderAalst-2016}. This new setting introduces several challenges: behavioural patterns must be detected in mined structures rather than imposed a priori; the input data may be incomplete or noisy; and the resulting logical specifications must be validated against real-world process behaviour. In response, we propose an updated pipeline and conduct a comprehensive empirical study using both illustrative and realistic logs, with a focus on validating structural and requirement-oriented properties through automated theorem proving. The method was independently re-implemented, and the experiments were conducted using newly developed tooling and data, reinforcing the generalisability and robustness of the approach.

To guide our investigation, we formulated the following research questions:
\begin{itemize}
    \item \textbf{RQ1}: What behavioural patterns can be reliably extracted from real-world event logs using workflow mining techniques?
    \item \textbf{RQ2}: To what extent do automatically generated logical specifications derived from mined workflows satisfy fundamental logical properties, such as satisfiability, consistency, and requirement compliance?
    \item \textbf{RQ3}: How does the level of noise present in event logs affect the structure, complexity, and correctness of the resulting logical specifications?
\end{itemize}


\section{Background and related work}

The automated extraction of formal models from workflow executions has been widely studied in both the process mining and formal methods communities. This section reviews key contributions related to process discovery, the generation of logical specifications, and the use of formal reasoning tools. We also highlight how our approach differs from and extends prior work in these areas.

\subsection{Workflow mining and process discovery}

Workflow mining aims to extract interpretable process models from recorded event logs, enabling the discovery of control-flow structures and behavioural dependencies based on observed executions.
Over the past two decades, various algorithms have been proposed to identify control-flow structures and behavioural patterns that underlie observed executions~\cite{vanderAalst-2016,Augusto-etal-2019}.
This general approach has also been recognised in broader reviews of
process model generation methods,
where the analysis of execution traces is highlighted as
a key technique for discovering temporal relationships between
activities~\cite{Wisniewski-etal-2019}.

Among the most prominent discovery algorithms are the Alpha Miner~\cite{vanderAalst-2004}, which identifies directly-follows relations, the Heuristics Miner~\cite{Weijters-etal-2006}, which improves robustness to noise via frequency-based rules, and the Inductive Miner~\cite{Leemans-etal-2013}, which recursively constructs block-structured models as process trees. The latter is particularly valued for its ability to produce sound and hierarchically structured models, even from noisy or incomplete logs~\cite{Augusto-etal-2019}.

Because our approach relies on process trees as an intermediate representation for generating logical specifications, we adopt the Inductive Miner as the primary discovery algorithm. Its ability to produce block-structured and semantically sound models makes it particularly suitable for our pattern-driven logic generation method.
Details concerning workflow mining are provided in Section~\ref{sec:mining-algorithms}.
Recent studies have also explored how low-level sensor data can be semantically abstracted into high-level event logs
suitable for process mining,
using techniques such as large language models to support industrial
process analysis~\cite{Brzychczy-etal-2024}.

\subsection{Process translation and formal modelling}

A substantial body of work has addressed the transformation of process models between various formal representations, particularly within business process management, to enable interoperability, formal analysis, or automated execution.
Established approaches include translations between Petri nets, process algebra, statecharts, UML activity diagrams, BPEL, and other executable
formalisms~\cite{Mendling-etal-2008,Ouyang-etal-2006b,Recker-Mendling-2006}.
These transformations typically aim to improve interoperability, enable simulation, or facilitate execution.

The concept of transforming behavioural models is not novel in computer science and
also encompasses process models.
These concepts are continually refined and updated,
garnering considerable interest and thereby imparting a sense of freshness to the field.
An exemplar of this is the translation between graphs and
web services~\cite{Mendling-etal-2008,vanderAalst-BisgaardLassen-2008}.
Process trees are frequently used in process discovery due to their hierarchical nature, which makes the structure of the model transparent from root to leaf and facilitates analysis and optimisation~\cite{vanZelst-Leemans-2020}.
Among the available discovery algorithms, the Inductive Miner has become the standard method for extracting process trees from event logs~\cite{Leemans-etal-2013}.

While many of these approaches preserve control-flow semantics, they rarely generate formal logical specifications that are directly suitable for automated reasoning. Efforts to bridge this gap include translating models into temporal logics (e.g., LTL, CTL) or modal logics for property verification~\cite{Ferilli-2016,Roubtsova-2005}. However, these methods often require manual intervention or assume domain-specific templates.

In contrast, our method does not translate graphical or executable models into other process languages, but derives logic-based specifications directly from mined process trees using a pattern-driven approach.

\subsection{Logic and automated reasoning}

Temporal logic has long served as a foundation for formalising system behaviour over time, particularly in the specification and verification of reactive and concurrent
systems~\cite{Emerson-1990,Manna-Pnueli-1992}. Linear temporal logic (LTL) and its variants allow properties such as safety, liveness, and fairness to be expressed in a declarative form and checked against system executions.

In the context of workflow analysis, several approaches have explored the use of logic-based formalisms to reason about process behaviour.
Early contributions translated either process models or execution traces into logical formalisms, most notably LTL or first-order logic (FOL), to express behavioural constraints and verify compliance~\cite{vanderAalst-etal-2005,Roubtsova-2005,Ferilli-2016}.
These approaches typically operate by encoding activity sequences as traces and then checking logical properties against those traces.

While prior approaches,
such as the work by van der Aalst et al.~\cite{vanderAalst-etal-2005},
define and test logical properties directly over event traces, our method differs by first mining a structured behavioural model and then generating a compositional logical specification subjected to formal reasoning using automated theorem provers.

Our method does not evaluate properties directly over execution traces. Instead, we mine a structured behavioural model in the form of a process tree and then automatically generate a compositional logical specification. This specification is analysed using automated theorem proving, avoiding the need for trace encoding or manual logic construction. As such, our approach bridges the gap between model discovery and formal verification.

\subsection{Automated logical specification generation}

Previous work~\cite{Klimek-2019-LAMP} demonstrated that logical specifications can be systematically derived from high-level process notations, such as UML activity diagrams, in order to support formal verification and reasoning.
This approach introduced a pattern-based translation framework, mapping behavioural fragments to reusable temporal logic templates.

In this study, we extend that line of research by shifting from manually defined models to data-driven structures extracted from event logs.
The framework now incorporates workflow mining as a front-end component, using process trees as an intermediate structure for logic generation. The resulting specifications are then analysed using automated reasoning tools, such as theorem provers, to verify properties such as satisfiability, logical implication, and requirement compliance.
This extension moves the technique toward empirical application and demonstrates its feasibility in settings where the process model is not predefined but discovered from real-world execution data.

This article extends our earlier work~\cite{Klimek-Witek-2024-ASE-RENE}
by incorporating two real-world event logs from healthcare and financial domains,
enabling a broader and more realistic empirical validation.
The evaluation is complemented by newly defined logical patterns,
including an extended conjunction operator $And4$,
and supported by a more detailed discussion of the results and their implications.
This version also introduces a dedicated analysis of threats to validity and expands
the experimental design to include multiple noise levels.
Overall, the article presents a more comprehensive and practically grounded assessment of
the proposed approach.

\section{Methodology}
\label{sec:methodology}

\subsection{Workflow mining algorithms}
\label{sec:mining-algorithms}

An \emph{event log} is a chronologically ordered record of activities observed within an information system. Event logs are widely used for purposes such as monitoring, auditing, debugging, and behavioural analysis~\cite{vanderAalst-2016}, and typically include event types, timestamps, resource identifiers, and process-related metadata. Workflow mining refers to the process of extracting behavioural models from such logs, with the aim of discovering process structures and understanding their execution dynamics.

A number of established algorithms are available for process discovery. Among the most prominent are:

\begin{itemize}
\item \textbf{Alpha Miner}~\cite{vanderAalst-2004,vanderAalst-2016,Augusto-etal-2019} is a foundational technique that identifies causal relations between activities based on directly-follows semantics. It constructs a process model by incrementally analysing observed successions of activities. Although conceptually simple and limited in handling complex behaviour (e.g., concurrency and noise), Alpha Miner remains a canonical baseline in the field.

\item \textbf{Heuristics Miner}~\cite{Weijters-etal-2006,vanderAalst-2016,Augusto-etal-2019} extends the Alpha Miner by introducing statistical thresholds and frequency-based heuristics. It accounts for noise and partial ordering by estimating the strength of dependencies between activities. This makes it better suited for real-world logs, which often contain exceptional behaviour or incomplete traces.

\item \textbf{Inductive Miner}~\cite{Leemans-etal-2013,vanderAalst-2016,Augusto-etal-2019} applies a recursive divide-and-conquer approach to discover structured, block-based process models. It guarantees soundness and produces models in the form of process trees, which makes it particularly well suited to formal transformation and automated reasoning. The algorithm is capable of handling concurrency, noise, and infrequent behaviour by adjusting a noise threshold parameter.
\end{itemize}

While these algorithms serve different purposes and apply various strategies, we focus on the Inductive Miner due to its structural guarantees and its ability to produce process trees-hierarchical models with explicit control-flow constructs. These trees serve as direct input to our pattern-based specification framework. In contrast, models generated by other algorithms, such as Petri nets or directly-follows graphs, would require additional transformation steps that may not preserve structural clarity or compositional semantics.

Workflow mining algorithms typically produce models in various formats, including Petri nets, process trees, or transition graphs. However, our approach is predicated on the use of process trees due to their well-defined hierarchical structure. Each process tree corresponds to a nested composition of control-flow operators (e.g., sequence, choice, concurrency), and can be mapped directly to predefined behavioural patterns~\cite{Szabo-2012,Recanti-2012,Tripakis-2016,Klimek-2019-LAMP}. This structural alignment enables the systematic generation of logical specifications in a compositional fashion, as described in Section~\ref{sec:generating-specification}.

Formally, a \emph{process tree} is a hierarchical structure in which internal nodes represent control-flow operators, while leaf nodes denote activities or silent transitions. It provides a well-formed abstraction that corresponds to a sound workflow net with guaranteed behavioural semantics. While the transformation from a process tree to a sound workflow net is always possible, the inverse transformation is non-trivial and requires sophisticated techniques, such as those proposed by van Zelst and Leemans~\cite{vanZelst-Leemans-2020}.

Among the algorithms reviewed, the Inductive Miner is particularly notable for producing sound models in a block-structured form by design. It has been shown to perform consistently well across a variety of benchmark datasets~\cite{Leemans-etal-2013,Bogarn-etal-2018}, and its process-tree output aligns naturally with our method's requirements.

In the context of our experiments, the application of the Inductive Miner to various event logs yielded a rich variety of pattern structures. Most discovered trees consisted of two to five activities per pattern instance, as illustrated in Formula~(\ref{for:approved-workflow}) and Definition~\ref{def:approved-workflow-syntax}. For exclusive choices and parallel constructs, the most common groupings included two or three distinct activities. The resulting structures occasionally involved loops and silent transitions, the latter introduced to ensure structural completeness. The set of approved patterns, defined in accordance with~\cite{Klimek-2019-LAMP}, remains fixed throughout the analysis but can be extended when necessary.

Certain patterns require the inclusion of designated entry ($s$) and exit ($e$) points, which may be mapped to \emph{null activities}. These artificial elements do not correspond to observable events and do not represent actual system actions. Their purpose is to maintain structural uniformity, facilitate translation, and preserve the semantic completeness of the model. Such activities have no duration or side effects and are added selectively during process tree construction.

\begin{remark}
The proposed method assumes that the mined process structure can be expressed as a process tree. While this format covers a broad class of practical workflows, unstructured or cyclic process models may require preprocessing or transformation to ensure compatibility with the pattern-based generation framework.
\end{remark}

\subsection{Logical specification generation}
\label{sec:generating-specification}

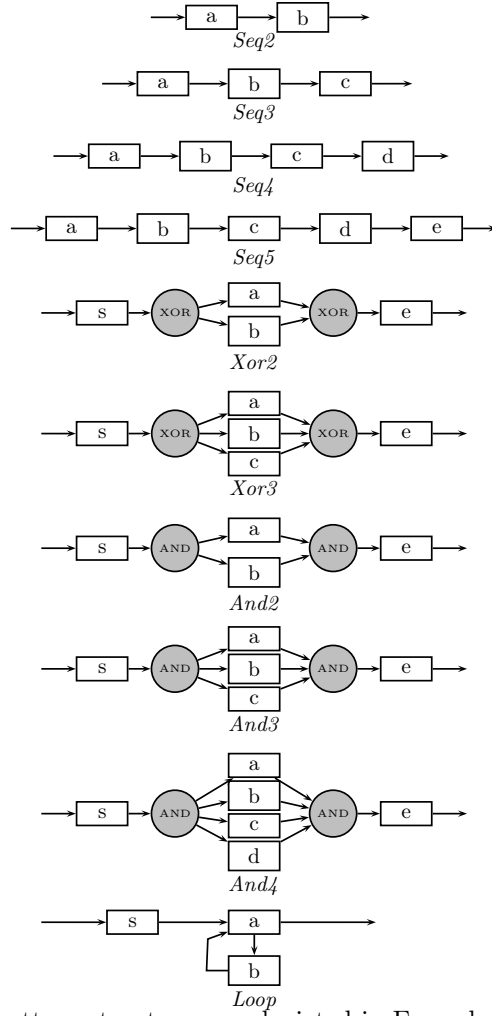
\begin{figure}[htb]
\centering
\scalebox{.8}{
\begin{pspicture}(8,16.5) 
      \psset{linecolor=black}

      \rput(2.3,16.3){\pnode{p2-0}{}}
      \rput(3.3,16.3){\rnode{p2-1}{\psframebox{\makebox[.6cm][c]{a}}}}
      \rput(4.8,16.3){\rnode{p2-2}{\psframebox{\makebox[.6cm][c]{b}}}}
      \rput(5.9,16.3){\pnode{p2-6}{}}
      \ncline{->}{p2-0}{p2-1}
      \ncline{->}{p2-1}{p2-2}
      \ncline{->}{p2-2}{p2-6}
      \rput(4,15.9){\normalsize \emph{Seq2}}

      \rput(1.5,15.2){\pnode{p3-0}{}}
      \rput(2.5,15.2){\rnode{p3-1}{\psframebox{\makebox[.6cm][c]{a}}}}
      \rput(4,15.2){\rnode{p3-2}{\psframebox{\makebox[.6cm][c]{b}}}}
      \rput(5.5,15.2){\rnode{p3-3}{\psframebox{\makebox[.6cm][c]{c}}}}
      \rput(6.6,15.2){\pnode{p3-6}{}}
      \ncline{->}{p3-0}{p3-1}
      \ncline{->}{p3-1}{p3-2}
      \ncline{->}{p3-2}{p3-3}
      \ncline{->}{p3-3}{p3-6}
      \rput(4,14.7){\normalsize \emph{Seq3}}

      \rput(.7,14.0){\pnode{p4-0}{}}
      \rput(1.7,14.0){\rnode{p4-1}{\psframebox{\makebox[.6cm][c]{a}}}}
      \rput(3.2,14.0){\rnode{p4-2}{\psframebox{\makebox[.6cm][c]{b}}}}
      \rput(4.7,14.0){\rnode{p4-3}{\psframebox{\makebox[.6cm][c]{c}}}}
      \rput(6.2,14.0){\rnode{p4-4}{\psframebox{\makebox[.6cm][c]{d}}}}
      \rput(7.2,14.0){\pnode{p4-6}{}}
      \ncline{->}{p4-0}{p4-1}
      \ncline{->}{p4-1}{p4-2}
      \ncline{->}{p4-2}{p4-3}
      \ncline{->}{p4-3}{p4-4}
      \ncline{->}{p4-4}{p4-6}
      \rput(4,13.5){\normalsize \emph{Seq4}}

      \rput(0,12.8){\pnode{p5-0}{}}
      \rput(1,12.8){\rnode{p5-1}{\psframebox{\makebox[.6cm][c]{a}}}}
      \rput(2.5,12.8){\rnode{p5-2}{\psframebox{\makebox[.6cm][c]{b}}}}
      \rput(4,12.8){\rnode{p5-3}{\psframebox{\makebox[.6cm][c]{c}}}}
      \rput(5.5,12.8){\rnode{p5-4}{\psframebox{\makebox[.6cm][c]{d}}}}
      \rput(7,12.8){\rnode{p5-5}{\psframebox{\makebox[.6cm][c]{e}}}}
      \rput(8,12.8){\pnode{p5-6}{}}
      \ncline{->}{p5-0}{p5-1}
      \ncline{->}{p5-1}{p5-2}
      \ncline{->}{p5-2}{p5-3}
      \ncline{->}{p5-3}{p5-4}
      \ncline{->}{p5-4}{p5-5}
      \ncline{->}{p5-5}{p5-6}
      \rput(4,12.3){\normalsize \emph{Seq5}}

      \rput(.5,11.4){\pnode{p5-0}{}}
      \rput(1.5,11.4){\rnode{p5-1}{\psframebox{\makebox[.6cm][c]{s}}}}
      \cnodeput[fillstyle=solid,fillcolor=lightgray](2.7,11.4){p5-2}{{\tiny XOR}}
      \rput(4,11.7){\rnode{p5-3}{\psframebox{\makebox[.6cm][c]{a}}}}
      \rput(4,11.1){\rnode{p5-5}{\psframebox{\makebox[.6cm][c]{b}}}}
      \cnodeput[fillstyle=solid,fillcolor=lightgray](5.3,11.4){p5-6}{{\tiny XOR}}
      \rput(6.5,11.4){\rnode{p5-7}{\psframebox{\makebox[.6cm][c]{e}}}}
      \rput(7.5,11.4){\pnode{p5-8}{}}
      \ncline{->}{p5-0}{p5-1}
      \ncline{->}{p5-1}{p5-2}
      \ncline{->}{p5-2}{p5-3}
      \ncline{->}{p5-2}{p5-5}
      \ncline{->}{p5-3}{p5-6}
      \ncline{->}{p5-5}{p5-6}
      \ncline{->}{p5-6}{p5-7}
      \ncline{->}{p5-7}{p5-8}
      \rput(4,10.6){\normalsize \emph{Xor2}}

      \rput(.5,9.4){\pnode{p6-0}{}}
      \rput(1.5,9.4){\rnode{p6-1}{\psframebox{\makebox[.6cm][c]{s}}}}
      \cnodeput[fillstyle=solid,fillcolor=lightgray](2.7,9.4){p6-2}{{\tiny XOR}}
      \rput(4,9.9){\rnode{p6-3}{\psframebox{\makebox[.6cm][c]{a}}}}
      \rput(4,9.4){\rnode{p6-4}{\psframebox{\makebox[.6cm][c]{b}}}}
      \rput(4,8.9){\rnode{p6-5}{\psframebox{\makebox[.6cm][c]{c}}}}
      \cnodeput[fillstyle=solid,fillcolor=lightgray](5.3,9.4){p6-6}{{\tiny XOR}}
      \rput(6.5,9.4){\rnode{p6-7}{\psframebox{\makebox[.6cm][c]{e}}}}
      \rput(7.5,9.4){\pnode{p6-8}{}}
      \ncline{->}{p6-0}{p6-1}
      \ncline{->}{p6-1}{p6-2}
      \ncline{->}{p6-2}{p6-3}
      \ncline{->}{p6-2}{p6-4}
      \ncline{->}{p6-2}{p6-5}
      \ncline{->}{p6-3}{p6-6}
      \ncline{->}{p6-4}{p6-6}
      \ncline{->}{p6-5}{p6-6}
      \ncline{->}{p6-6}{p6-7}
      \ncline{->}{p6-7}{p6-8}
      \rput(4,8.5){\normalsize \emph{Xor3}}

      \rput(.5,7.5){\pnode{p7-0}{}}
      \rput(1.5,7.5){\rnode{p7-1}{\psframebox{\makebox[.6cm][c]{s}}}}
      \cnodeput[fillstyle=solid,fillcolor=lightgray](2.7,7.5){p7-2}{{\tiny AND}}
      \rput(4,7.8){\rnode{p7-3}{\psframebox{\makebox[.6cm][c]{a}}}}
      \rput(4,7.1){\rnode{p7-5}{\psframebox{\makebox[.6cm][c]{b}}}}
      \cnodeput[fillstyle=solid,fillcolor=lightgray](5.3,7.5){p7-6}{{\tiny AND}}
      \rput(6.5,7.5){\rnode{p7-7}{\psframebox{\makebox[.6cm][c]{e}}}}
      \rput(7.5,7.5){\pnode{p7-8}{}}
      \ncline{->}{p7-0}{p7-1}
      \ncline{->}{p7-1}{p7-2}
      \ncline{->}{p7-2}{p7-3}
      \ncline{->}{p7-2}{p7-5}
      \ncline{->}{p7-3}{p7-6}
      \ncline{->}{p7-5}{p7-6}
      \ncline{->}{p7-6}{p7-7}
      \ncline{->}{p7-7}{p7-8}
      \rput(4,6.6){\normalsize \emph{And2}}

      \rput(.5,5.5){\pnode{p8-0}{}}
      \rput(1.5,5.5){\rnode{p8-1}{\psframebox{\makebox[.6cm][c]{s}}}}
      \cnodeput[fillstyle=solid,fillcolor=lightgray](2.7,5.5){p8-2}{{\tiny AND}}
      \rput(4,6.0){\rnode{p8-3}{\psframebox{\makebox[.6cm][c]{a}}}}
      \rput(4,5.5){\rnode{p8-4}{\psframebox{\makebox[.6cm][c]{b}}}}
      \rput(4,5.0){\rnode{p8-5}{\psframebox{\makebox[.6cm][c]{c}}}}
      \cnodeput[fillstyle=solid,fillcolor=lightgray](5.3,5.5){p8-6}{{\tiny AND}}
      \rput(6.5,5.5){\rnode{p8-7}{\psframebox{\makebox[.6cm][c]{e}}}}
      \rput(7.5,5.5){\pnode{p8-8}{}}
      \ncline{->}{p8-0}{p8-1}
      \ncline{->}{p8-1}{p8-2}
      \ncline{->}{p8-2}{p8-3}
      \ncline{->}{p8-2}{p8-4}
      \ncline{->}{p8-2}{p8-5}
      \ncline{->}{p8-3}{p8-6}
      \ncline{->}{p8-4}{p8-6}
      \ncline{->}{p8-5}{p8-6}
      \ncline{->}{p8-6}{p8-7}
      \ncline{->}{p8-7}{p8-8}
      \rput(4,4.6){\normalsize \emph{And3}}

      \rput(.5,3.1){\pnode{p9-0}{}}
      \rput(1.5,3.1){\rnode{p9-1}{\psframebox{\makebox[.6cm][c]{s}}}}
      \cnodeput[fillstyle=solid,fillcolor=lightgray](2.7,3.1){p9-2}{{\tiny AND}}
      \rput(4,3.9){\rnode{p9-3}{\psframebox{\makebox[.6cm][c]{a}}}}
      \rput(4,3.4){\rnode{p9-4}{\psframebox{\makebox[.6cm][c]{b}}}}
      \rput(4,2.9){\rnode{p9-5}{\psframebox{\makebox[.6cm][c]{c}}}}
      \rput(4,2.4){\rnode{p9-6}{\psframebox{\makebox[.6cm][c]{d}}}}
      \cnodeput[fillstyle=solid,fillcolor=lightgray](5.3,3.1){p9-7}{{\tiny AND}}
      \rput(6.5,3.1){\rnode{p9-8}{\psframebox{\makebox[.6cm][c]{e}}}}
      \rput(7.5,3.1){\pnode{p9-9}{}}
      \ncline{->}{p9-0}{p9-1}
      \ncline{->}{p9-1}{p9-2}
      \ncline{->}{p9-2}{p9-3}
      \ncline{->}{p9-2}{p9-4}
      \ncline{->}{p9-2}{p9-5}
      \ncline{->}{p9-2}{p9-6}
      \ncline{->}{p9-3}{p9-7}
      \ncline{->}{p9-4}{p9-7}
      \ncline{->}{p9-5}{p9-7}
      \ncline{->}{p9-6}{p9-7}
      \ncline{->}{p9-7}{p9-8}
      \ncline{->}{p9-8}{p9-9}
      \rput(4,1.9){\normalsize \emph{And4}}

      \rput(.5,1.3){\pnode{p10-0}{}}
      \rput(2,1.3){\rnode{p10-1}{\psframebox{\makebox[.6cm][c]{s}}}}
      \rput(4,1.3){\rnode{p10-2}{\psframebox{\makebox[.6cm][c]{a}}}}
      \rput(4,.5){\rnode{p10-3}{\psframebox{\makebox[.6cm][c]{b}}}}
      \rput(6,1.3){\pnode{p10-4}{}}
      \ncline{->}{p10-0}{p10-1}
      \ncline{->}{p10-1}{p10-2}
      \ncline{->}{p10-2}{p10-3}
      \ncdiag[angleA=180,angleB=200]{->}{p10-3}{p10-2}
      \ncline{->}{p10-2}{p10-4}
      \rput(4,0){\normalsize \emph{Loop}}
\end{pspicture}
}
\caption{Approved pattern structures,
as depicted in Formula~(\ref{for:approved-workflow}) and Definition~\ref{def:approved-workflow-syntax}}
\label{fig:approved-workflow-semantics}
\end{figure}

As we are considering a new application area,
namely event logs and workflow mining,
it is necessary to introduce new predefined behavioural patterns compared to those introduced in
article~\cite{Klimek-2019-LAMP}.
Our experiments with the Inductive Miner indicate that the following set of approved patterns is sufficient:
\begin{eqnarray}
\Pi = \{ Seq2, Seq3, Seq4, Seq5, Xor2, Xor3, And2, And3, And4, Loop\} \label{for:approved-workflow}
\end{eqnarray}
They define the various types of activity sequences,
exclusive choice of activities,
parallel execution of activities,
and loops with activities.
Definition~\ref{def:approved-workflow-syntax} and
Figure~\ref{fig:approved-workflow-semantics}
illustrate their syntax and semantics.
Compared to the pattern set introduced in~\cite{Klimek-Witek-2024-ASE-RENE},
these definitions have been extended to incorporate
the newly introduced $And4$ pattern,
which emerged from the analysis of the two real-world event logs discussed in
Section~\ref{sec:empirical-study-design},
where this construct proved to be of practical relevance.
This extension also highlights the adaptability of
the underlying specification generation framework,
which remains open to integrating additional predefined behavioural patterns as required.

\begin{definition}
\label{def:approved-workflow-syntax}
The approved set of patterns is defined as follows:
\begin{itemize}
\item $Seq2(a,b)\equiv a;b$
\item $Seq3(a,b,c)\equiv a;b;c$
\item $Seq4(a,b,c,d)\equiv a;b;c;d$
\item $Seq5(a,b,c,d,e)\equiv a;b;c;d;e$
\item $Xor2(s,a,b,e)\equiv s;(a\otimes b);e$
\item $Xor3(s,a,b,c,e)\equiv s;(a\otimes b\otimes c);e$
\item $And2(s,a,b,e)\equiv s;(a\|b);e$
\item $And3(s,a,b,c,e)\equiv s;(a\|b\|c);e$
\item $And4(s,a,b,c,d,e)\equiv s;(a\|b\|c\|d);e$
\item $Loop(s,a,b)\equiv s; a \mbox{~~~or~~~} s;a;b;a \mbox{~~~or~~~} s;a;b;a;b;a \mbox{~~~etc.}$
\end{itemize}
where $s$, $e$, $a$, $b$, $c$ and $d$ represent (atomic) activities, that is pattern formal arguments.
';' operator denotes sequential execution,
'$\|$' denotes concurrent execution,
and '$\otimes$' denotes exclusive choice execution for specified activities.
The whole expression is always processed from left to right,
and parentheses enforce the joint consideration of arguments.
(The definition of the last pattern is inductive in nature.)
\end{definition}
It should be noted that the introduced patterns are partially redundant.
For instance,
instead of using multiple patterns such as $Seq2$-$5$,
one could limit the approach to a single pattern $Seq$ with only two arguments.
This redundancy is intentional and is incorporated for experimental and notational convenience.
\begin{fact}
The pattern set introduced in Formula~(\ref{for:approved-workflow}) and
Definition~\ref{def:approved-workflow-syntax},
and illustrated in Figure~\ref{fig:approved-workflow-semantics},
is supported by empirical evidence obtained in this study across four event logs,
including two real-world cases.
The results confirm that the fixed set effectively captures
the most frequently occurring behavioural patterns.
Section~\ref{sec:empirical-study-design} provides detailed justification for
the completeness and practical applicability of the pattern set.
\end{fact}
\begin{remark}
The $Loop$ pattern is also referred to as $Xor$\,$Loop$, as illustrated in
Figures~\ref{fig:process-tree-runningexample}--\ref{fig:process-tree-repairexample-noise-1}.
Null activities, commonly denoted by ``$tau$'', are introduced to complete pattern structure when needed;
see Formulas~(\ref{for:w1})--(\ref{for:w2}).
The argument ``$s$'' in the $Loop$ pattern is always assigned a null activity to ensure correct structural form.
\end{remark}

\emph{Null activities} are artificially introduced elements,
if necessary, with no operational counterpart or execution value.
They maintain model structural consistency and facilitate analysis,
completing the required pattern structure.
Null activities are also sometimes referred to as \emph{silent activities}.
\begin{remark}
The use of silent activities in workflow discovery is widely recognised and well justified in the literature~\cite{Wen-etal-2010,Leemans-etal-2024}.
\end{remark}

\begin{table}[htb]
\centering
\caption{A set of fixed logical properties in Propositional Linear Temporal Logic for approved patterns}
{\small
\begin{tabularx}{1\columnwidth}{|cXl|}
\hline
& &\\
$\Sigma=\{$ &
$\mathbf{Seq2}(a,b)=
\langle
a,
b,
\som a,
\alw(a \imp \som b),
\alw\neg(a \con b)
\rangle,$
&\\
&
$\mathbf{Seq3}(a,b,c)=
\langle
a,
c,
\som a,
\alw(a \imp \som b),
\alw(b \imp \som c),
\alw\neg(a \con b),
\alw\neg(a \con c),
\alw\neg(b \con c)
\rangle,$
&\\
&
$\mathbf{Seq4}(a,b,c,d)=
\langle
a,
d,
\som a,
\alw(a \imp \som b),
\alw(b \imp \som c),
\alw(c \imp \som d),
\alw\neg(a \con b),
\alw\neg(a \con c),
\alw\neg(a \con d),
\alw\neg(b \con c),
\alw\neg(b \con d),
\alw\neg(c \con d)
\rangle,$
&\\
&
$\mathbf{Seq5}(a,b,c,s,e)=
\langle
a,
e,
\som a,
\alw(a \imp \som b),
\alw(b \imp \som c),
\alw(c \imp \som d),
\alw(d \imp \som e),
\alw\neg(a \con b),
\alw\neg(a \con c),
\alw\neg(a \con d),
\alw\neg(a \con e),
\alw\neg(b \con c),
\alw\neg(b \con d),
\alw\neg(b \con e),
\alw\neg(c \con d),
\alw\neg(c \con e),
\alw\neg(d \con e)
\rangle,$
&\\
&
$\mathbf{Xor2}(s,a,b,e)=
\langle
s,
e,
\som s,
\alw(s \imp (\som a \con \neg\som b) \dis (\neg\som a \con \som b)),
\alw((a \dis b) \imp \som e),
\alw\neg(s \con a),
\alw\neg(s \con b),
\alw\neg(s \con e),
\alw\neg(a \con b),
\alw\neg(a \con e),
\alw\neg(b \con e)
\rangle,$
&\\
&
$\mathbf{Xor3}(s,a,b,c,e)=
\langle
s,
e,
\som s,
\alw(s \imp (\som a \con \neg\som b \con \neg\som c) \dis (\neg\som a \con \som b \con \neg\som c) \dis (\neg\som a \con \neg\som b \con \som c)),
\alw((a \dis b \dis c) \imp \som e),
\alw\neg(s \con a),
\alw\neg(s \con b),
\alw\neg(s \con c),
\alw\neg(s \con e),
\alw\neg(a \con b),
\alw\neg(a \con c),
\alw\neg(a \con e),
\alw\neg(b \con c),
\alw\neg(b \con e),
\alw\neg(c \con e)
\rangle,$
&\\
&
$\mathbf{And2}(s,a,b,e)=
\langle
s,
e,
\som s,
\alw(s \imp \som a \con \som b),
\alw(a \imp \som e),
\alw(b \imp \som e),
\alw\neg(s \con (a \dis b)),
\alw\neg((a \dis b) \con e)
\rangle,$
&\\
&
$\mathbf{And3}(s,a,b,c,e)=
\langle
s,
e,
\som s,
\alw(s \imp \som a \con \som b \con \som c),
\alw(a \imp \som e),
\alw(b \imp \som e),
\alw(c \imp \som e),
\alw\neg(s \con (a \dis b \dis c)),
\alw\neg((a \dis b \dis c) \con e)
\rangle,$
&\\
&
$\mathbf{And4}(s,a,b,c,d,e)=
\langle
s,
e,
\som s,
\alw(s \imp \som a \con \som b \con \som c \con \som d),
\alw(a \imp \som e),
\alw(b \imp \som e),
\alw(c \imp \som e),
\alw(d \imp \som e),
\alw\neg(s \con (a \dis b \dis c \dis d)),
\alw\neg((a \dis b \dis c \dis d) \con e)
\rangle,$
&\\
&
$\mathbf{Loop}(s,a,b)=
\langle
s,
a,
\som s,
\alw(s \imp \som a),
\alw(a \imp (\som b \con \som a) \dis \neg\som b),
\alw(b \imp \som a),
\alw\neg(s \con a),
\alw\neg(s \con b),
\alw\neg(a \con b)
\rangle$
$\}$
& \\
& & \\
\hline
\end{tabularx}
}
\label{tab:approved-workflow-PLTL}
\end{table}

\begin{figure*}[!htb]
\def\database{\scalebox{.6}{\pspicture(0,-0.2)(2,1.1)
  \psset{linecolor=black!70}
  \pscustom[fillstyle=solid,fillcolor=gray!30]{%
    \psline(0,0)(0,1)
    \pscurve(1,1.1)(2,1)
    \psline(2,0)
    \pscurve(1,-0.1)(0,0)}%
    \pscurve(0,1)(1,0.9)(2,1)
    \pscurve(0,0.9)(1,0.8)(2,0.9)
    \pscurve(0,0.8)(1,0.7)(2,0.8)
  \endpspicture}}
\centering
\scalebox{.9}
{
\begin{pspicture}(16,3) 
\rput(.5,.7){\rnode{pkt:dia}{\database}}
\rput(3.4,.7){\rnode{pkt:miner}{\psframebox
                      {\begin{tabular}{c}
                            \textcolor{black}{Miner}\\
                            \textcolor{black}{\textit{tree builder}}
                      \end{tabular}}}}
\rput(6.5,.7){\rnode{pkt:pattern}{\psframebox
                      {\begin{tabular}{c}
                            \textcolor{black}{Pattern}\\
                            \textcolor{black}{\textit{extractor}}
                      \end{tabular}}}}
\rput(6.5,2.5){\rnode{pkt:pi}{\psframebox
                      {\begin{tabular}{c}
                            \textcolor{black}{\textit{define} $\Pi$}
                      \end{tabular}}}}
\rput(9.5,.7){\rnode{pkt:gen}{\psframebox
                      {\begin{tabular}{c}
                            \textcolor{black}{Logical spec.}\\
                            \textcolor{black}{\textit{generator}}
                      \end{tabular}}}}
\rput(9.5,2.5){\rnode{pkt:sigma}{\psframebox
                      {\begin{tabular}{c}
                            \textcolor{black}{\textit{define} $\Sigma$}
                      \end{tabular}}}}
\rput(12.8,.7){\rnode{pkt:alr}{\psframebox
                      {\begin{tabular}{c}
                            \textcolor{black}{Logical engine}\\
                            \textcolor{black}{\textit{reasoner}}
                      \end{tabular}}}}
\rput(12.8,2.5){\rnode{pkt:req}{\psframebox
                      {\begin{tabular}{c}
                            \textcolor{black}{\textit{define} $Rq$}
                      \end{tabular}}}}
\rput(15.8,.7){\rnode{pkt:log}{\database}}
\ncline{->}{pkt:dia}{pkt:miner}\Aput{\small Event}\Bput{\small log}
\ncline{->}{pkt:miner}{pkt:pattern}\Aput{\small $Tree$}
\ncline{->}{pkt:pi}{pkt:pattern}\Aput{\small $\Pi$}
\ncline{->}{pkt:pattern}{pkt:gen}\Aput{\small $W$}
\ncline{->}{pkt:sigma}{pkt:gen}\Aput{\small $\Sigma$}
\ncline{->}{pkt:gen}{pkt:alr}\Aput{\small $L$}
\ncline{->}{pkt:req}{pkt:alr}\Aput{\small $Rq$}
\ncline{->}{pkt:alr}{pkt:log}\Aput{\small Hints}\Bput{\begin{tabular}{c}\small Sys\\ logs\end{tabular}}
\end{pspicture}
}
\caption{General scheme of data processing and flows}
\label{fig:architecture}
\end{figure*}
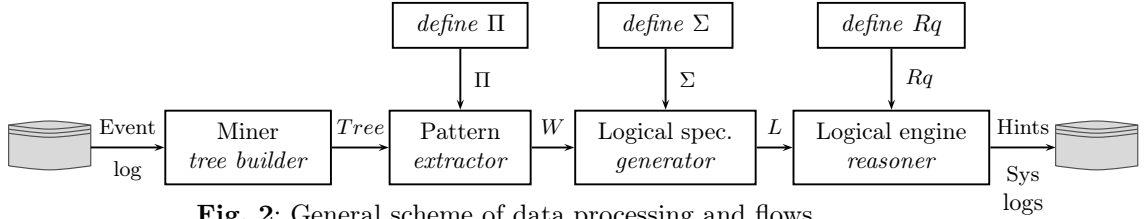

The logical patterns for
Formula~(\ref{for:approved-workflow})
are presented in
Table \ref{tab:approved-workflow-PLTL}.
Defining these logical patterns requires some expertise,
but once defined,
they can be repeatedly used,
even by inexperienced users, see processing in
Figure~\ref{fig:architecture}.
This method ensures both satisfiability preservation and
relative completeness~\cite{Klimek-2019-LAMP,Klimek-2014-AMCS}.

The entire processing pipeline involves two steps:
\begin{enumerate}
\item
Initially
a pattern expression $W$
as a literal representation of the tree is constructed:
\begin{eqnarray}
\Pi extract(Tree,\Pi) = W\label{for:Pi-ext}
\end{eqnarray}
Here, $Tree$ represents the tree as a result of workflow mining,
while $\Pi$ denotes the set of approved patterns.
\item
Algorithm $\Pi C$~\cite{Klimek-2019-LAMP} generates the logical specification:
\begin{eqnarray}
\Pi C(W,\Sigma) = L\label{for:Pi-C}
\end{eqnarray}
where $\Sigma$ signifies the logical definitions of patterns,
and $L$ is the resulting logical specification.
\end{enumerate}

Figure~\ref{fig:architecture}
summarizes the system architecture and key data flows.
The event log is retrieved from the database,
analysed by a workflow miner (Section~\ref{sec:mining-algorithms}),
and an event tree is generated.
This tree is scanned,
and a pattern expression $W$ is constructed using approved processing patterns $\Pi$,
representing the process tree.
A logical specification $L$, equivalent to the workflow,
is then automatically generated based on logical patterns $\Sigma$ and
tested by theorem provers for logical satisfiability or other logical relations with added requirements $Rq$.
The results are stored in the database.
Patterns $\Pi$ and $\Sigma$ are defined once and reused, while properties $Rq$ are defined as needed.

\section{Empirical study design}
\label{sec:empirical-study-design}

Workflow extraction primarily uses
the Inductive Miner algorithm.
Logical patterns,
defined in Propositional Linear-Time Temporal Logic (PLTL)~\cite{Emerson-1990,Wolter-Wooldridge-2011},
are tested using First-order Logic (FOL)~\cite{Kleene-1952} theorem provers due to the broader availability of FOL engines.
Although PLTL is more expressive~\cite{Emerson-1990,vanBenthem-1995},
FOL is used as a minimal temporal logic~\cite{vanBenthem-1995}.
This simplification,
substituting $\alw$ and $\som$ operators with universal and existential quantifiers,
respectively,
allows for the immediate use of available FOL provers for experiments.
Future experiments will reconsider using PLTL provers.

We have chosen the following two theorem provers
for the experiments conducted in this study:
E~\cite{E-tool,Schulz-2002} and
Vampire~\cite{Vampire-tool,Riazanov-Voronkov-2002} for FOL.
The selection is justified by their reputation and effectiveness, particularly due to several factors, namely:
effectiveness, accuracy and active community support.
\begin{itemize}
\item
effectiveness --
both Vampire and E are well-known in the academic and industrial environments for
their effectiveness in solving satisfiability-related problems;
\item
accuracy --
both tools are continuously developed and tested, ensuring high accuracy of results.
Their algorithms are based on advanced mathematical and computational techniques.
\item
active community support --
both tools benefit from the support of active communities of users and developers,
providing access to documentation, technical support,
and the ability to report and resolve issues.
\end{itemize}
Both provers require input data in the \emph{TPTP} format~\cite{Sutcliffe-2017},
to which specifications are accordingly translated.
Upon execution,
they return a \texttt{SZS status} --
a standardised outcome code used by automated theorem provers to classify
the result of a proof attempt.
It indicates whether the conjecture has been proven,
refuted, or remains unresolved,
based on its logical consistency with the provided axioms.


\begin{remark}
Theorem provers such as Vampire and E return
an \texttt{SZS status},
a standardised outcome code indicating the result of
a proof attempt with respect to a given set of axioms and a conjecture.
\texttt{SZS status Theorem} or \texttt{SZS status Unsatisfiable}
indicates that the conjecture logically follows from the axioms,
typically established by refuting its negation.
\texttt{SZS status CounterSatisfiable}
means that the axioms together with the negated conjecture are satisfiable,
thus, a counterexample exists in which the conjecture does not hold.
\texttt{SZS status ContradictoryAxioms}
signals inconsistency within the axioms themselves, independent of the conjecture.
These classifications provide a uniform semantic framework for interpreting prover results~\cite{Schreiner-2023}.
A summary is provided in Table~\ref{tab:prover-results}.
\end{remark}

\begin{table}[!htb]
\centering
\caption{Typical \texttt{SZS status} outputs of provers and their interpretation}
\label{tab:prover-results}
\begin{tabular}{ll}
\toprule
\textbf{SZS status} & \textbf{Interpretation} \\
\midrule
\texttt{Unsatisfiable} / \texttt{Theorem} & Property holds (negation refuted) \\
\texttt{Satisfiable} & Conjecture not provable (negation consistent) \\
\texttt{CounterSatisfiable} & Conjecture falsified (counterexample exists) \\
\texttt{ContradictoryAxioms} & Axioms are internally inconsistent \\
\texttt{NoProofFound} / timeout & No result -- prover inconclusive \\
\bottomrule
\end{tabular}
\end{table}

\subsection{Data sources}

The following event logs were considered in the experimental evaluation:
\begin{enumerate}
\item
\textbf{\texttt{running-example.xes}}:
a synthetic log representing a general-purpose administrative process.
This log serves as a minimal test case for
verifying the correctness of pattern extraction and logical specification
generation\footnote{Available at \url{https://pm4py.fit.fraunhofer.de/getting-started-page} (previously accessible on 18.09.2023)
and additionally available at \url{https://github.com/process-intelligence-solutions/pm4py/blob/release/tests/input_data/running-example.xes} (accessed on 20.05.2025)}.
\item
\textbf{\texttt{repair-example.xes}}:
a synthetic log reflecting a generic service or complaint-handling process,
incorporating conditional paths and optional repair-related
activities\footnote{Available at \url{https://ai.ia.agh.edu.pl/pl:dydaktyka:dss:lab02} (previously accessible on 18.09.2023)
and additionally available at \url{http://home.agh.edu.pl/~kluza/repairExample.xes} (accessed on 20.05.2025)}.
\item
\textbf{Hospital Billing}:
a real-world event log obtained from a hospital billing system, capturing both clinical and administrative procedures related to patient
discharge\footnote{Available at \url{https://data.4tu.nl/articles/_/12705113/1} (last accessed on 20.05.2025)}.
\item
\textbf{BPI Challenge 2012}:
a real-world event log recording the lifecycle of loan applications, comprising key process steps such as submission, assessment, approval, and eventual
rejection\footnote{Available at \url{https://data.4tu.nl/articles/_/12689204/1} (last accessed on 20.05.2025)}.
\end{enumerate}
The first two event logs, while not artificially generated, represent general-purpose processes designed primarily for illustrative or educational use. In contrast, the latter two logs originate from the recording of real-world cases and reflect actual operational workflows in healthcare and financial services,
see Table~\ref{tab:log-comparison}.

\begin{table}[ht]
\centering
\caption{Comparison of event logs used in the study}
\label{tab:log-comparison}
\begin{tabular}{@{}llll@{}}
\toprule
\textbf{Event Log} & \textbf{Origin} & \textbf{Domain} & \textbf{Purpose} \\ \midrule
\texttt{running-example.xes} & General-purpose (non-synthetic) & Abstract / administrative & Educational, demonstrative \\
\texttt{repair-example.xes}  & General-purpose (non-synthetic) & Service / repair handling & Educational, demonstrative \\
\texttt{Hospital Billing}    & Real-world process data        & Healthcare & Empirical benchmarking \\
\texttt{BPI Challenge 2012}  & Real-world process data        & Financial services & Empirical benchmarking \\
\bottomrule
\end{tabular}
\end{table}

\begin{figure}[htb]
    \centering
    \includegraphics[width=0.55\textwidth]{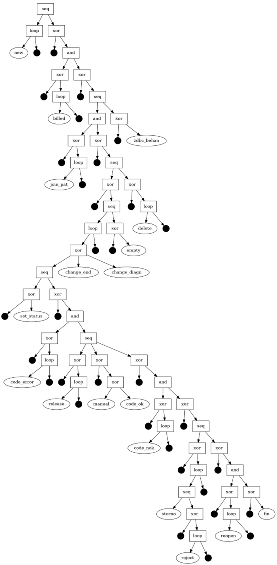}
    \caption{Process tree obtained from the real-world \texttt{Hospital Billing} event log
    with a noise parameter value of 0}
\label{fig:process-tree-hospital-billing}
\end{figure}

\begin{figure}[htb]
    \centering
    \includegraphics[width=1.1\textwidth]{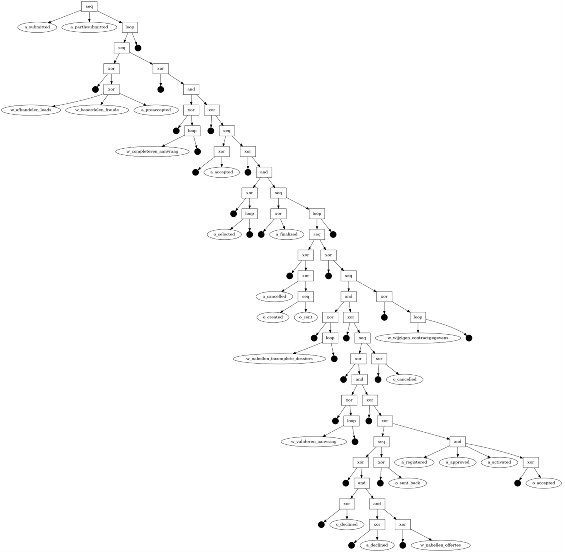}
    \caption{Process tree obtained from the real-world \texttt{BPI Challenge 2012} event log
    with a noise parameter value of 0}
\label{fig:process-tree-bpic}
\end{figure}

The corresponding process trees derived from these logs are presented in
Figures~\ref{fig:process-tree-runningexample},
\ref{fig:process-tree-repairexample},
\ref{fig:process-tree-hospital-billing}, and
\ref{fig:process-tree-bpic}, respectively.
All process trees were generated using the Inductive Miner algorithm.
Silent (null) transitions,
denoted as $\tau$, are visible in
Figure~\ref{fig:process-tree-repairexample} and were introduced to
preserve the structural integrity of the mined patterns.
(In the two subsequent trees,
such silent transitions are also present and appear visually as black dots.)

\begin{figure}[htb]
\centering
\scalebox{.7}
{
\pstree[levelsep=7ex]{\Toval{Seq}}
{
\TR{Register\_request}
\pstree{\Toval{Xor loop}}{
                          \pstree{\Toval{Seq}}{
                             \pstree{\Toval{And}}{
                                \TR{Check\_ticket}
                                \pstree{\Toval{Xor}}{\TR{Examine\_thoroughly}\TR{Examine\_casually}}
                                }
                             \TR{Decide}
                                              }
                          \TR{Reinitiate\_request}
                          }
\pstree{\Toval{Xor}}{\TR{Reject\_request}\TR{Pay\_compensation}}
}
}
\caption{Process tree for the event log \texttt{running-example.xes}}
\label{fig:process-tree-runningexample}
\end{figure}

\begin{figure}[htb]
\centering
\scalebox{.7}
{
\pstree[levelsep=7ex]{\Toval{Seq}}
{
\TR{Register}
\TR{Analyse\_defect}
\pstree{\Toval{And}}{
        \pstree{\Toval{Xor}}{\TR{tau}\TR{Inform\_user}}
        \pstree{\Toval{And}}{
                 \pstree{\Toval{Xor}}{
                         \TR{tau}
                         \pstree{\Toval{Xor loop}}{{\TR{Test\_repair}}{\TR{tau}}}
                                     }
                 \pstree{\Toval{Xor loop}}{\pstree{\Toval{Xor}}{\TR{Repair\_simple}\TR{Repair\_complex}}\TR{Restart\_repair}}
                            }
                    }
\pstree{\Toval{Xor}}{\TR{tau}\TR{Archive\_repair}}
\TR{End}
}
}
\caption{Process tree for the event log \texttt{repair-example.xes}}
\label{fig:process-tree-repairexample}
\end{figure}

There are pattern expressions presented below for
Figures~\ref{fig:process-tree-runningexample}
and~\ref{fig:process-tree-repairexample},
respectively:
\begin{eqnarray}
W_1 = Seq3(Register\_request,Loop(Seq2(And2\nonumber\\
(Check\_ticket,Xor2(Examine\_thoroughly,\nonumber\\
Examine\_casually)),Decide),Reinitiate\_request),\nonumber\\
Xor2(Reject\_request,Pay\_compensation))\label{for:w1}
\end{eqnarray}
\begin{eqnarray}
W_2 = Seq5(Register,Analyse\_defect,And2\nonumber\\
(Xor2(tau1,Inform\_user),And2(Xor2(tau2,\nonumber\\
Loop(Test\_repair,tau3)),Loop(Xor2\nonumber\\
(Repair\_complex,Repair\_simple),Restart\_repair))),\nonumber\\
Xor2(tau4,Archive\_repair),End)\label{for:w2}
\end{eqnarray}
They constitute a literal representation of process trees,
while simultaneously serving as output data, see~Formula~(\ref{for:Pi-ext}),
and input data, see~Formula~(\ref{for:Pi-C}),
in the processing pipeline, see Figure~\ref{fig:architecture}.

\begin{figure}[htb]
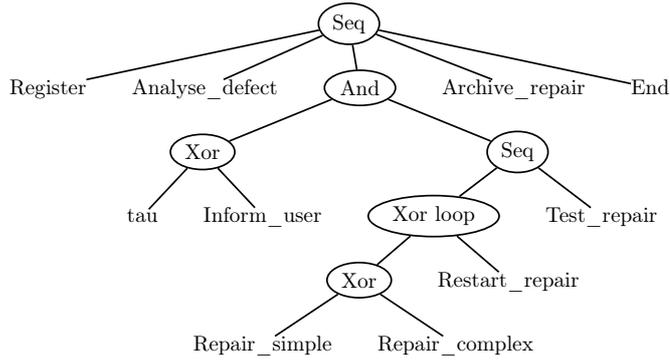

\centering
\scalebox{.8}
{
\pstree[levelsep=7ex]{\Toval{Seq}}
{
\TR{Register}
\TR{Analyse\_defect}
\pstree{\Toval{And}}{
  \pstree{\Toval{Xor}}{\TR{tau}\TR{Inform\_user}}
  \pstree{\Toval{Seq}}{
         \pstree{\Toval{Xor loop}}{
                \pstree{\Toval{Xor}}{\TR{Repair\_simple}\TR{Repair\_complex}}
                \TR{Restart\_repair}
                                  }
         \TR{Test\_repair}
                      }
}
\TR{Archive\_repair}
\TR{End}
}
}
\caption{Process tree for the event log \texttt{repair-example.xes} with a noise parameter value of 0.5}
\label{fig:process-tree-repairexample-noise-0-5}
\end{figure}

\begin{figure}[htb]
    \centering
    \includegraphics[width=1\textwidth]{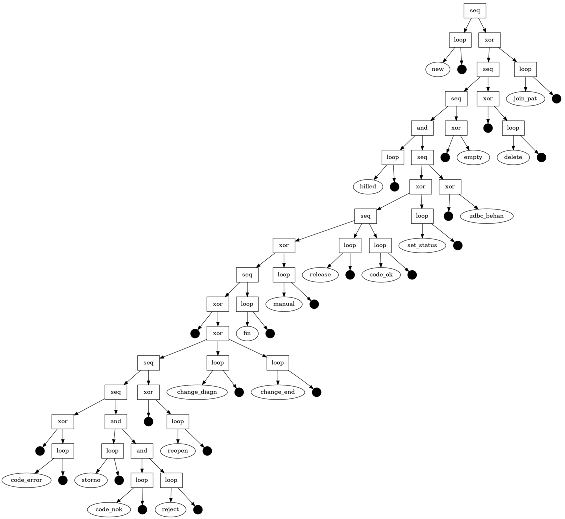}
    \caption{Process tree obtained from the real-world \texttt{Hospital Billing} event log
    with a noise parameter value of 0.25}
\label{fig:process-tree-hospital-billing-noise-0-25}
\end{figure}

\begin{figure}[htb]
    \centering
    \includegraphics[width=1.05\textwidth]{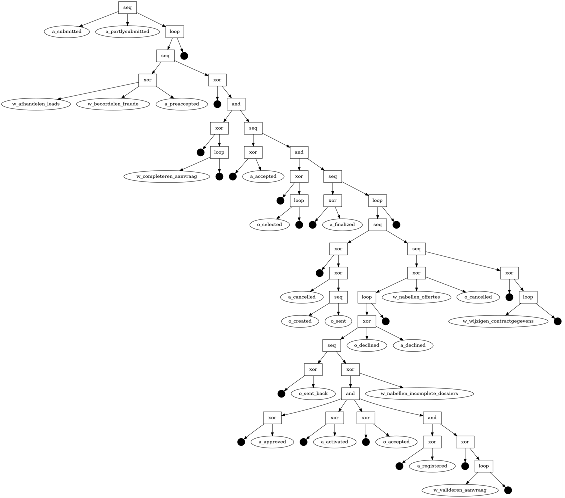}
    \caption{Process tree obtained from the real-world \texttt{BPI Chalenge 2012} event log
    with a noise parameter value of 0.25}
\label{fig:process-tree-bpic-noise-0-25}
\end{figure}

The Inductive Miner algorithm supports configurable noise thresholds,
allowing for the extraction of process models that are either
highly detailed or more generalised,
depending on the level of data uncertainty.
By adjusting the noise parameter,
we are able to explore how incomplete or noisy event logs influence
the structure and complexity of the resulting process trees.
This capability is particularly valuable for empirical investigation,
as it enables the controlled simulation of real-world imperfections in input data.
Figures~\ref{fig:process-tree-repairexample-noise-0-5}--\ref{fig:process-tree-bpic-noise-0-25}
illustrate process trees derived from event logs
each processed with a different noise parameter setting.

\begin{figure}[htb]
\centering
\scalebox{.8}
{
\pstree[levelsep=7ex]{\Toval{Seq}}
{
\TR{Register}
\TR{Analyse\_defect}
\pstree{\Toval{And}}{
  \TR{Inform\_user}
  \pstree{\Toval{Xor}}{
         \TR{tau}
         \pstree{\Toval{Xor loop}}{\TR{Test\_repair}\TR{tau}}
         \TR{Repair\_complex}
                      }
}
\TR{Archive\_repair}
\TR{End}
}
}
\caption{Process tree for the event log \texttt{repair-example.xes} with a noise parameter value of 1}
\label{fig:process-tree-repairexample-noise-1}
\end{figure}

\subsection{Experimental setup}
\label{sec:results-analysis}


This evaluation is based on four event logs,
covering both illustrative and real-world scenarios.
The experimental pipeline follows the methodology introduced in Section~\ref{sec:methodology},
consisting of mining, pattern mapping, specification generation, and property validation.
The selection of logs and parameter settings enables systematic testing across various conditions,
including noise levels and requirement types.

\begin{remark}
The conducted experiments are categorised into three groups:
\begin{enumerate}
\item
testing the satisfiability of the obtained logical specifications --
this process assesses the logical soundness of the derived formulas;
\item
testing the logical relations between the obtained logical specifications --
this method evaluates the logical consistency and interrelations among the selected sets of formulas;
\item
testing the fulfilment of logical properties, expressed by new and separate formulas,
with respect to the obtained logical specifications --
this process introduces new requirement formulas and evaluates their relationship with those automatically derived from event logs.
\end{enumerate}
\end{remark}
This three-pronged approach ensures that the experiments comprehensively address the key aspects of logical analysis,
ranging from consistency to correctness achievability,
thereby creating a robust framework for evaluating the various aspects of the obtained logical specifications.

\subsection{Satisfiability testing}

Logical specification and satisfiability testing is crucial for ensuring the basic system correctness,
reducing the risk of errors in later project stages.
The following experimental questions are formulated,
represented as a logical formula in a separate file:
\begin{itemize}
\item
\textbf{Problem1.p} --
the logical problem generated for the event log
\texttt{running-example.xes}
at the default noise threshold level,
see Figure~\ref{fig:process-tree-runningexample};
\item
\textbf{Problem2.p} --
the logical problem generated for the event log
\texttt{repair-example.xes}
at the default noise threshold level,
see Figure~\ref{fig:process-tree-repairexample};
\item
\textbf{Problem3.p} --
the logical problem generated for the event log
\texttt{repair-example.xes}
with the noise threshold set to 0.5,
see Figure~\ref{fig:process-tree-repairexample-noise-0-5};
\item
\textbf{Problem4.p} --
the logical problem generated for the event log
\texttt{repair-example.xes}
with the noise threshold set to 1,
see Figure~\ref{fig:process-tree-repairexample-noise-1}.
\end{itemize}
\emph{Noise} refers to the presence of random, unwanted or
irrelevant activities that can affect the analysis of business processes.
Many workflow mining algorithms allow to set a noise parameter in order to filter rare behaviours out,
thus influencing the accuracy of the entire process of building the behavioural model.
Table~\ref{tab:problem-1-4} summarizes the inference results for particular problems and provers.
%
\begin{fact}
Testing the satisfiability of a logical specification,
see Table~\ref{tab:problem-1-4},
derived from a process model enables the verification of its fundamental consistency. This provides a basic yet essential validation step of the underlying axioms, ensuring that the specification admits at least one admissible model of behaviour.
\end{fact}

\begin{table}[htb]
\caption{Results of testing logical Problems~1--4 for both provers,
or the four basic logical specifications obtained}
\label{tab:problem-1-4}
\centering
{\small
\begin{tabularx}{\columnwidth}{l|X|X}
\toprule
File & Vampire's output & E's output \\
\cmidrule{2-3}
 & Comment & Comment \\
\toprule
\multirow{ 2}{*}{Problem1.p} &
\texttt{\%~SZS status Satisfiable for problem1 \%~Termination reason: Satisfiable} & \texttt{\#~No proof found! \#~SZS staus Satisfiable}\\
\cmidrule{2-3}
  & Specification satisfied & Specification satisfied\\
\hline
\multirow{ 2}{*}{Problem2.p} &
\texttt{\%~SZS status Satisfiable for tptp \%~Termination reason: Satisfiable} & \texttt{\#~No proof found! \#~SZS staus Satisfiable}\\
\cmidrule{2-3}
  & Specification satisfied & Specification satisfied\\
\hline
\multirow{ 2}{*}{Problem3.p} &
\texttt{\%~SZS status Satisfiable for problem3 \%~Termination reason: Satisfiable} & \texttt{\#~No proof found! \#~SZS status Satisfiable}\\
\cmidrule{2-3}
  & Specification satisfied & Specification satisfied\\
\hline
\multirow{ 2}{*}{Problem4.p} &
\texttt{\%~SZS status Satisfiable for problem4 \%~Termination reason: Satisfiable} & \texttt{\#~No proof found! \#~SZS status Satisfiable}\\
\cmidrule{2-3}
  & Specification satisfied & Specification satisfied\\
\bottomrule
\end{tabularx}
}
\end{table}

\subsection{Logical relationships}

The analysis may also include examining logical relationships between the available specifications.
These relationships are vital for ensuring coherence, modularity,
and facilitating system verification and understanding.
We considered the following issues:
\begin{itemize}
\item
\textbf{Problem5.p} -- the question is whether the specification for Problem~2 logically entails the specification for Problem~3;
\item
\textbf{Problem6.p} -- the question is whether the specification for Problem~3 logically entails the specification for Problem~4;
\item
\textbf{Problem7.p} -- the question is whether the specification for Problem~1 logically entails the specification for Problem~4;
\item
\textbf{Problem8.p} --
the question is whether the specification for Problem~2 is logically equivalent to the specification for Problem~3;
\item
\textbf{Problem9.p} --
the question is whether the specification for Problem~3 is logically equivalent to the specification for Problem~4;
\item
\textbf{Problem10.p} --
the question is whether the specification for Problem~1 is logically equivalent to the specification for Problem~4.
\end{itemize}
Examining relationships are crucial for ensuring system correctness,
understanding its structure and interdependencies, managing complexity,
and facilitating modifications,
especially when varying the levels of workflow extraction detail.
Other relationships may also be explored.
Table~\ref{tab:problem-5-10} presents the inference results for specific problems and provers.
\begin{fact}
Testing logical relationships between derived specifications allows for assessing their coherence and structural correspondence.
The results in Table~\ref{tab:problem-5-10}
show that all the problems are logically resolved.
However, the outcome also depends on how the axioms and the conjecture are formulated,
as even minor changes may affect the prover's result.
\end{fact}

\begin{table}[!htb]
\caption{Results of testing logical Problems~5--10, or the logical relationships between basic specifications}
\label{tab:problem-5-10}
\centering
{\small
\begin{tabularx}{\columnwidth}{l|X|X}
\toprule
File & Vampire's output & E's output \\
\cmidrule{2-3}
   & Comment & Comment \\
\toprule
\multirow{ 2}{*}{Problem5.p} &
\texttt{\%~SZS status CounterSatisfiable for tpt \%~Termination reason: Satisfiable} & \texttt{\#~No proof found! \#~SZS status CounterSatisfiable}\\
\cmidrule{2-3}
  & Satisfiable axioms, refuted conjecture & Satisfiable axioms, refuted conjecture\\
\hline
\multirow{ 2}{*}{Problem6.p} &
\texttt{\%~SZS status CounterSatisfiable for problem6 \%~Termination reason: Satisfiable} & \texttt{\#~No proof found! \#~SZS status CounterSatisfiable}\\
\cmidrule{2-3}
  & Satisfiable axioms, refuted conjecture & Satisfiable axioms, refuted conjecture\\
\hline
\multirow{ 2}{*}{Problem7.p} &
\texttt{\%~SZS status CounterSatisfiable for problem7 \%~Termination reason: Satisfiable} & \texttt{\#~No proof found! \#~SZS status CounterSatisfiable}\\
\cmidrule{2-3}
  & Satisfiable axioms, refuted conjecture & Satisfiable axioms, refuted conjecture\\
\hline
\multirow{ 2}{*}{Problem8.p} &
\texttt{\%~SZS status CounterSatisfiable for tpt \%~Termination reason: Satisfiable} & \texttt{\#~No proof found! \#~SZS status CounterSatisfiable}\\
\cmidrule{2-3}
  & Satisfiable axioms, refuted conjecture & Satisfiable axioms, refuted conjecture\\
\hline
\multirow{ 2}{*}{Problem9.p} &
\texttt{\%~SZS status CounterSatisfiable for problem9 \%~Termination reason: Satisfiable} & \texttt{\#~No proof found! \#~SZS status CounterSatisfiable}\\
\cmidrule{2-3}
  & Satisfiable axioms, refuted conjecture & Satisfiable axioms, refuted conjecture\\
\hline
\multirow{ 2}{*}{Problem10.p} &
\texttt{\%~SZS status CounterSatisfiable for problem10 \%~Termination reason: Satisfiable} & \texttt{\#~No proof found! \#~SZS status CounterSatisfiable}\\
\cmidrule{2-3}
  & Satisfiable axioms, refuted conjecture & Satisfiable axioms, refuted conjecture\\
\bottomrule
\end{tabularx}
}
\end{table}

\subsection{Logical requirements}

Studying logical specifications in relation to the requirement-specific logical formulas ensures compliance,
identifies discrepancies and guarantees system correctness.
It also facilitates requirement traceability from high-level to low-level.
It is crucial to examine two categories of formulas:
\emph{liveness properties}~\cite{Lamport-1977,Alpern-Schneider-1985}
and \emph{safety properties}~\cite{Lamport-1977,Alpern-Schneider-1985}.
Any set of logical formulas for the system must include both types of
properties~\cite{Kindler-1994,Alpern-Schneider-1985,Manna-Pnueli-1992}.
These properties can be verified separately or together in relation to the model and specification.

For the specification of Problem~1, two requirements have been defined, namely liveness:
\begin{eqnarray}
\alw(Register\_request \imp
\som(Reject\_request \dis Pay\_compensation))\label{for:problem1-liveness}
\end{eqnarray}
and safety:
\begin{eqnarray}
\alw\neg(Reject\_request \con Pay\_compensation)\label{for:problem1-safety}
\end{eqnarray}
In turn,
for the specification of Problems~3 and~4, two requirements have been defined, that is liveness:
\begin{eqnarray}
\alw(Register \imp \som(Repair\_simple \dis Repair\_complex))\label{for:problem3-4-liveness}
\end{eqnarray}
and safety:
\begin{eqnarray}
\alw\neg(Inform\_user \con null)\label{for:problem3-4-safety}
\end{eqnarray}
We expect these properties to be fulfilled.

Thus, the last group of experiments involves verifying whether
particular specifications $L$ satisfy certain logical properties $r$ that we expect.
Utilizing the \emph{deduction theorem}~\cite{Kleene-1952,Ben-Ari-2012},
logical formula $\vdash L \imp r$
constitutes the input for the selected logical engines,
that is we test whether particular specification implies properties:
\begin{itemize}
\item
\textbf{Problem11.p} --
the question is whether the specification for Problem~1 implies Formula~(\ref{for:problem1-liveness});
\item
\textbf{Problem12.p} --
the question is whether the specification for Problem~1 implies Formula~(\ref{for:problem1-safety});
\item
\textbf{Problem13.p} --
the question is whether the specification for Problem~3 implies Formula~(\ref{for:problem3-4-liveness});
\item
\textbf{Problem14.p} --
the question is whether the specification for Problem~3 implies Formula~(\ref{for:problem3-4-safety});
\item
\textbf{Problem15.p} --
the question is whether the specification for Problem~4 implies Formula~(\ref{for:problem3-4-liveness});
\item
\textbf{Problem16.p} --
the question is whether the specification for Problem~4 implies Formula~(\ref{for:problem3-4-safety}).
\end{itemize}
Table~\ref{tab:problem-11-16} presents the inference results for particular problems and provers.

\begin{table}[!htb]
\caption{Results of testing logical Problems~11--16,
or basic specifications against workflow correctness requirements}
\label{tab:problem-11-16}
\centering
{\small
\begin{tabularx}{\columnwidth}{l|X|X}
\toprule
File & Vampire's output & E's output \\
\cmidrule{2-3}
   & Comment & Comment \\
\toprule
\multirow{ 2}{*}{\begin{tabular}{ll}Problem11.p\\\mbox{}\\ \textit{liveness formula}\end{tabular}} &
\texttt{\%~SZS status Theorem for problem11 \%~Termination reason: Satisfiable} & \texttt{\#~Proof found! \#~SZS status Theorem}\\
\cmidrule{2-3}
  & Theorem established & Theorem established\\
\hline
\multirow{ 2}{*}{\begin{tabular}{ll}Problem12.p\\\mbox{}\\ \textit{safety formula}\end{tabular}} &
\texttt{\%~SZS status Theorem for problem12 \%~Termination reason: Satisfiable} & \texttt{\#~Proof found! \#~SZS status Theorem}\\
\cmidrule{2-3}
  & Theorem established & Theorem established\\
\hline
\multirow{ 2}{*}{\begin{tabular}{ll}Problem13.p\\\mbox{}\\ \textit{liveness formula}\end{tabular}} &
\texttt{\%~SZS status Theorem for problem13 \%~Termination reason: Satisfiable} & \texttt{\#~Proof found! \#~SZS status Theorem}\\
\cmidrule{2-3}
  & Theorem established & Theorem established\\
\hline
\multirow{ 2}{*}{\begin{tabular}{ll}Problem14.p\\\mbox{}\\ \textit{safety formula}\end{tabular}} &
\texttt{\%~SZS status Theorem for problem14 \%~Termination reason: Satisfiable} & \texttt{\#~Proof found! \#~SZS status Theorem}\\
\cmidrule{2-3}
  & Theorem established & Theorem established\\
\hline
\multirow{ 2}{*}{\begin{tabular}{ll}Problem15.p\\\mbox{}\\ \textit{liveness formula}\end{tabular}} &
\texttt{\%~SZS status CounterSatifsiable for problem15 \%~Termination reason: Refutation} & \texttt{\#~No proof found! \#~SZS status CounterSatisfiable}\\
\cmidrule{2-3}
  & Conjecture falsified & Conjecture falsified\\
\hline
\multirow{ 2}{*}{\begin{tabular}{ll}Problem16.p\\\mbox{}\\ \textit{safety formula}\end{tabular}} &
\texttt{\%~SZS status Theorem for problem16 \%~Termination reason: Satisfiable} & \texttt{\#~Proof found! \#~SZS status Theorem}\\
\cmidrule{2-3}
  & Theorem established & Theorem established\\
\bottomrule
\end{tabularx}
}
\end{table}

\begin{fact}
Verifying specifications against domain-specific logical requirements ensures that
the extracted process models reflect intended behaviour.
The results in Table~\ref{tab:problem-11-16}
demonstrate that all formulas combined with the desired properties are satisfied.
However, when the noise level increases during workflow mining, certain behavioural guarantees may no longer hold, leading to unsatisfied liveness conditions or unmet assumptions. This reflects the natural degradation of logical completeness in less structured or noisy data.
\end{fact}

\begin{fact}
The proposed approach consistently supports the full pipeline,
from event logs and workflow mining to the generation and logical validation of formal specifications. All stages were executed without interruption, confirming the method's end-to-end feasibility in practice.
\end{fact}

\subsection{Real-world case study}

Building on the approach applied to the previously examined illustrative event logs, we extend our analysis to include two real-case scenarios. For each of these, we define a set of domain-specific logical requirements that serve as the basis for evaluating the generated logical specifications. These requirements enable the systematic assessment of key formal properties, such as satisfiability, consistency, and logical entailment. In particular, the real-world logs from the hospital billing system and the BPI Challenge 2012 dataset offer rich, complex process structures grounded in actual operational practice. Evaluating our method in these realistic contexts allows us to investigate whether the mined models preserve the behavioural constraints that are expected in domains such as healthcare and financial services.

For the first hospital case,
we examine the satisfiability of domain-relevant properties, including the following liveness requirements:
\begin{eqnarray}
\som(new \con \som join\_pat)\label{for:problem21-liveness}\\
\alw(new \imp \som join\_pat)\label{for:problem22-liveness}
\end{eqnarray}
We analyse two liveness properties linking $new$ and $join\_pat$ events.
The first formula
asserts that this sequence occurs at least once,
while the second requires it
after every $new$, ensuring consistent patient flow.

For the second financial case,
we examine the satisfiability of domain-specific constraints related to
application status and task coordination:
\begin{eqnarray}
\alw \neg(w\_afhandelen\_leads \wedge a\_preaccepted)\label{for:problem27-safety}\\
\alw(a\_submitted \imp \som a\_partly\_submitted)\label{for:problem28-liveness}
\end{eqnarray}
We analyse two properties:
the first enforces mutual exclusion (safety) between two activities that should not co-occur,
ensuring logical consistency in process execution.
The second expresses a liveness condition requiring that every submitted application is
eventually processed at least partially.

For the hospital case, we consider the following logical problems:
\begin{itemize}
\item
\textbf{Problem17.p} --
satisfiability for the logical problem generated for the event log
\textbf{Hospital Billing}
at the noise threshold set to 0,
see Figure~\ref{fig:process-tree-hospital-billing};
\item
\textbf{Problem18.p} --
satisfiability for the logical problem generated for the event log
\textbf{Hospital Billing}
at the noise threshold set to 0.25,
see Figure~\ref{fig:process-tree-hospital-billing-noise-0-25};
\item
\textbf{Problem19.p} --
the question is whether the specification for Problem~17 is logically equivalent to the specification for Problem~18;
\item
\textbf{Problem20.p} --
the question is whether the specification for Problem~17 logically entails the specification for Problem~18;
\item
\textbf{Problem21.p} --
the question is whether the specification for Problem~17 implies Formula~(\ref{for:problem21-liveness});
\item
\textbf{Problem22.p} --
the question is whether the specification for Problem~18 implies Formula~(\ref{for:problem22-liveness});
\end{itemize}

For the financial case, we consider the following logical problems:
\begin{itemize}
\item
\textbf{Problem23.p} --
satisfiability for the logical problem generated for the event log
\textbf{BPI Challenge 2012}
at the noise threshold set to 0,
see Figure~\ref{fig:process-tree-bpic};
\item
\textbf{Problem24.p} --
satisfiability for the logical problem generated for the event log
\textbf{BPI Challenge 2012}
at the noise threshold set to 0.25,
see Figure~\ref{fig:process-tree-bpic-noise-0-25};
\item
\textbf{Problem25.p} --
the question is whether the specification for Problem~23 logically entails the specification for Problem~24;
\item
\textbf{Problem26.p} --
the question is whether the specification for Problem~23 is logically equivalent to the specification for Problem~24;
\item
\textbf{Problem27.p} --
the question is whether the specification for Problem~23 implies Formula~(\ref{for:problem27-safety});
\item
\textbf{Problem28.p} --
the question is whether the specification for Problem~24 implies Formula~(\ref{for:problem28-liveness});
\end{itemize}

The results of testing these two groups of logical properties are summarised in
Tables~\ref{tab:problem-17-22}--\ref{tab:problem-23-28}, respectively.

\begin{table}[!htb]
\caption{Results of testing logical Problems~17--22 across all test categories}
\label{tab:problem-17-22}
\centering
{\small
\begin{tabularx}{\columnwidth}{l|X|X}
\toprule
File & Vampire's output & E's output \\
\cmidrule{2-3}
   & Comment & Comment \\
\toprule
\multirow{ 2}{*}{\begin{tabular}{ll}Problem17.p\\\mbox{}\\ \end{tabular}} &
\texttt{\%~SZS status Satisfiable for tptp \%~Termination reason: Satisfiable} & \texttt{\#~No proof found! \#~SZS status Satisfiable}\\
\cmidrule{2-3}
  & Specification satisfied & Specification satisfied\\
\hline
\multirow{ 2}{*}{\begin{tabular}{ll}Problem18.p\\\mbox{}\\ \end{tabular}} &
\texttt{\%~SZS status Unsatisfiable for tptp \%~Termination reason: Refutation} & \texttt{\#~Proof found! \#~SZS status Unsatisfiable}\\
\cmidrule{2-3}
  & Specification unsatisfied & Specification unsatisfied\\
\hline
\multirow{ 2}{*}{\begin{tabular}{ll}Problem19.p\\\mbox{}\\ \end{tabular}} &
\texttt{\%~SZS status Theorem for tptp\_EQUIVALENT \%~Termination reason: Refutation} &
\texttt{\#~Proof found! \#~SZS status ContradictoryAxioms}\\
\cmidrule{2-3}
  & The proof refutation & Contradictory between axioms\\
\hline
\multirow{ 2}{*}{\begin{tabular}{ll}Problem20.p\\\mbox{}\\ \end{tabular}} &
\texttt{\%~SZS status Theorem for tptp\_IMPLIES \%~Termination reason: Refutation} &
\texttt{\#~Proof found! \#~SZS status ContradictoryAxioms}\\
\cmidrule{2-3}
  & The proof refutation & Contradictory between axioms\\
\hline
\multirow{ 2}{*}{\begin{tabular}{ll}Problem21.p\\\mbox{}\\ \textit{liveness formula}\end{tabular}} &
\texttt{\%~SZS status Theorem for tptp\_thesis \%~Termination reason: Refutation} & \texttt{\#~Proof found! \#~SZS status Theorem}\\
\cmidrule{2-3}
  & Conjecture falsified & Conjecture falsified\\
\hline
\multirow{ 2}{*}{\begin{tabular}{ll}Problem22.p\\\mbox{}\\ \textit{liveness formula}\end{tabular}} &
\texttt{\%~SZS status Theorem for tptp\_thesis \%~Termination reason: Refutation} & \texttt{\#~Proof found! \#~SZS status Theorem}\\
\cmidrule{2-3}
  & Conjecture falsified & Conjecture falsified\\
\bottomrule
\end{tabularx}
}
\end{table}

\begin{table}[!htb]
\caption{Results of testing logical Problems~23--28 across all test categories}
\label{tab:problem-23-28}
\centering
{\small
\begin{tabularx}{\columnwidth}{l|X|X}
\toprule
File & Vampire's output & E's output \\
\cmidrule{2-3}
   & Comment & Comment \\
\toprule
\multirow{ 2}{*}{\begin{tabular}{ll}Problem23.p\\\mbox{}\\ \end{tabular}} &
\texttt{\%~SZS status Unsatisfiable for tptp \%~Termination reason: Refutation} & \texttt{\#~Proof found! \#~SZS status Unsatisfiable}\\
\cmidrule{2-3}
  & Specification unsatisfied & Specification unsatisfied\\
\hline
\multirow{ 2}{*}{\begin{tabular}{ll}Problem24.p\\\mbox{}\\ \end{tabular}} &
\texttt{\%~SZS status Satisfiable for tptp \%~Termination reason: Satisfiable} & \texttt{\#~No proof found! \#~SZS status Satisfiable}\\
\cmidrule{2-3}
  & Specification satisfied & Specification satisfied\\
\hline
\multirow{ 2}{*}{\begin{tabular}{ll}Problem25.p\\\mbox{}\\ \end{tabular}} &
\texttt{\%~SZS status Theorem for tptp\_IMPLIES \%~Termination reason: Refutation} &
\texttt{\#~Proof found! \#~SZS status ContradictoryAxioms}\\
\cmidrule{2-3}
  & The proof refutation & Contradictory between axioms\\
\hline
\multirow{ 2}{*}{\begin{tabular}{ll}Problem26.p\\\mbox{}\\ \end{tabular}} &
\texttt{\%~SZS status Theorem for tptp\_EQUIVALENT \%~Termination reason: Refutation} &
\texttt{\#~Proof found! \#~SZS status ContradictoryAxioms}\\
\cmidrule{2-3}
  & The proof refutation & Contradictory between axioms\\
\hline
\multirow{ 2}{*}{\begin{tabular}{ll}Problem27.p\\\mbox{}\\ \textit{safety formula}\end{tabular}} &
\texttt{\%~SZS status Theorem for tptp\_thesis \%~Termination reason: Refutation} & \texttt{\#~Proof found! \#~SZS status Theorem}\\
\cmidrule{2-3}
  & Conjecture falsified & Conjecture falsified\\
\hline
\multirow{ 2}{*}{\begin{tabular}{ll}Problem28.p\\\mbox{}\\ \textit{liveness formula}\end{tabular}} &
\texttt{\%~SZS status CounterSatisfiable for tptp\_thesis \%~Termination reason: Satisfiable} & \texttt{\#~No proof found! \#~SZS status CounterSatisfiable}\\
\cmidrule{2-3}
  & Theorem established & Theorem established\\
\bottomrule
\end{tabularx}
}
\end{table}

\begin{fact}
The validation of logical specifications derived from real-world event logs confirmed,
see Tables~\ref{tab:problem-17-22}-\ref{tab:problem-23-28},
that the proposed approach remains effective in preserving and
verifying domain-relevant behavioural constraints.
These results are
consistent with those observed for illustrative logs,
demonstrating the method's applicability beyond synthetic scenarios.
\end{fact}

\begin{fact}
The predefined pattern set proved sufficient for all analysed cases, yet the specification generation framework remains extensible. New patterns,
such as $And4$ introduced in this study,
can be added to accommodate evolving process structures without altering the core method.
\end{fact}

\section{Discussion}

The evaluation aimed to assess the feasibility of generating logic-based specifications from process logs and validating their properties using automated reasoning. The experiments were conducted across a range of process logs, including both illustrative examples and real-case scenarios. Results were organised around three research questions, focusing on pattern extraction, logical structure, and the impact of data quality. The findings provide insight into the practical scope and limitations of the approach.

With respect to RQ1, the experimental results show that all considered logs could be effectively translated into structured process trees using the Inductive Miner, and subsequently mapped onto the predefined workflow pattern set. This held true even for real-world logs with complex domain-specific behaviour, such as the Hospital Billing and BPI Challenge datasets. The pattern base proved expressive enough to capture common control-flow constructs, confirming the general applicability of the method to diverse sources of event data.

RQ2 concerned the generation of logical specifications and their alignment with key formal properties. Across all logs and tool versions, the derived specifications were systematically subjected to property-based testing, including satisfiability analysis, logical ralationship, and validation against domain-specific requirements. These tests provided evidence of internal coherence and allowed for a comparative analysis of different tool configurations and noise levels. The use of theorem provers enabled this validation to be performed in a formal, automated setting.

RQ3 addressed the effect of noise on the logical output. Introducing noise into the logs led to observable variation in the structural completeness of the mined process trees and, in turn, in the outcome of property validation. Some logical constraints became more difficult to verify under noisy conditions. Nonetheless, the use of extended pattern definitions and an improved implementation enhanced the method's resilience, as certain expected behaviours could still be identified and tested, even when the input data was partially degraded.

\begin{table}[!htb]
\centering
\caption{Summary of structural and logical metrics across four event logs}
\label{tab:event-log-summary}
\begin{tabularx}{\textwidth}{lrrrccccc}
\toprule
\textbf{Event log} &
\rotatebox{-90}{\thead{File\\size}} &
\rotatebox{-90}{\thead{Number of\\rows}} &
\rotatebox{-90}{\thead{Registered\\events}} &
\rotatebox{-90}{\thead{Discovered\\activities (all)}} &
\rotatebox{-90}{\thead{Activities (no $\tau$)}} &
\rotatebox{-90}{\thead{Used patterns}} &
\rotatebox{-90}{\thead{Process tree\\height}} &
\rotatebox{-90}{\thead{Generated\\formulas}} \\
\midrule
Running Example      & 15.8 KB  & 380       & 42       & 8  & 8  & 6  & 6  & 36  \\
Repair Example       & 6.55 MB  & 171\,345  & 17\,676  & 13 & 8  & 9  & 6  & 64  \\
Hospital Billing     & 170 KB   & 4\,487\,686 & 451\,359 & 51 & 18 & 48 & 26 & 354 \\
BPI Challenge 2012   & 72 KB    & 1\,633\,365 & 262\,200 & 53 & 24 & 49 & 26 & 361 \\
\bottomrule
\end{tabularx}
\end{table}

The comparative analysis of structural metrics, summarised in Table~\ref{tab:event-log-summary}, highlights how the proposed method performs across logs of varying scale and origin. The mined process trees reveal that even in high-volume logs, the number of distinct behavioural actions remains manageable, while the height and structure of the tree reflect the degree of nesting and internal regularity present in the processes.

The number of generated logical formulas corresponds to the structural richness of the mined tree. It increases with the number of applied patterns and the height of the model, rather than with surface-level attributes such as file size or total event count. This confirms the method's scalability and its robustness in handling both illustrative examples and large-scale empirical data.

Silent activities, denoted as $\tau$, arise naturally during process mining and serve to represent optional or skipped actions. In our approach, they may also be introduced deliberately to enforce syntactic alignment with predefined behavioural patterns. Such insertions are meaningful and structural, not incidental, and they support the compositional logic generation framework implemented in this study.

The file size or the number of log entries should not be interpreted as indicators of logical complexity. Factors such as repeated traces, long process instances, or verbose event encoding contribute to data volume but not necessarily to structural depth. Our aim is not to benchmark process mining algorithms themselves, but to build on the interpretable output produced by established techniques. In this regard, the Inductive Miner provides reliable process trees that serve as a stable basis for logical abstraction.

The formula counts refer to unfiltered tree models. Adjusting the noise threshold during mining would modify the extracted structure and reduce the number of resulting formulas. Furthermore, validation tasks involving comparisons between two specifications, such as logical implication or equivalence, naturally require a doubled formula base.

Solver runtimes were not recorded, as modern theorem provers handle specifications of this scale efficiently. Our work~\cite{Klimek-Semczyszyn-2025-EASE} has shown that significantly larger theories can be resolved with negligible latency, supporting the feasibility of integrating real-time logical validation into modelling environments.

Taken together, these findings confirm the method's practical viability. Its output scales with structural complexity rather than data volume, and the logical specifications it produces are readily amenable to automated reasoning and validation.


%
%
%

\section{Threats to validity}

We acknowledge several threats to the validity of our empirical study, categorised below according to standard classifications.

\paragraph{Internal validity.}
Threats to internal validity primarily concern the correctness of the implemented toolchain and the reliability of the transformation from mined process structures to logical specifications. The underlying algorithm for logical specification generation has been validated in earlier work~\cite{Klimek-2019-LAMP}, and the set of predefined behavioural patterns used in this study is sufficient to fully cover the structure of the mined process trees. All identified patterns were matched deterministically, and the trees generated by the Inductive Miner were confirmed to be structurally correct. While the transformation relies on rule-based mappings, no mismatches were observed in the recognition or application of patterns. In addition, potential variability introduced by automated theorem provers was controlled by applying consistent encoding schemes and solver configurations throughout the evaluation. Selected property checks were cross-validated using alternative provers to confirm alignment in outcomes.
Another threat is related to the formulation of axioms and conjectures. Even minor changes in the encoding of logical queries may affect the outcome of automated provers, especially in cases where semantics is only approximated due to the FOL simplification of PLTL constructs.

\paragraph{External validity.}
The evaluation includes both illustrative and real-world event logs, providing a representative basis for assessing the general applicability of the approach. The two real-case scenarios cover structurally rich and practically relevant processes, offering evidence that the method can handle realistic behavioural complexity. As the proposed framework operates on process trees-hierarchical and compositional structures-and the Inductive Miner algorithm is designed to produce such models directly, we do not anticipate fundamental barriers in applying the method to other domains, provided that logs meet basic quality and completeness conditions. Furthermore, the logical specification generation process is open to incorporating new behavioural patterns. The predefined pattern base can be extended with additional constructs as needed, as long as they preserve structural integrity, which supports the method's adaptability to a wide range of workflows and application contexts.

\paragraph{Construct validity.}
Construct validity concerns the extent to which the evaluation captures meaningful and appropriate aspects of the studied phenomenon. In this work, the selected logical properties--such as satisfiability, internal consistency, and implication relations--directly reflect the structural and behavioural correctness of generated specifications. These aspects are well aligned with the primary goal of the method, which is to support formal reasoning and identify potential inconsistencies within and across models. Although the current evaluation focuses on a core set of logical relations, the framework remains open to incorporating additional, more complex properties depending on domain needs.
Furthermore, we complemented logical validation with a set of structural metrics,
such as the number of discovered activities, applied workflow patterns, tree depth,
and the resulting size of the logical specification.
These quantitative characteristics enhance the interpretability of
the specifications and provide an additional perspective on their scope and complexity.

\paragraph{Conclusion validity.}
Conclusion validity relates to the extent to which the results support the stated claims. The conclusions in this study are derived from a structured evaluation of logical properties across multiple event logs and systematically varied experimental conditions. Observed behaviours and model responses were consistent with expectations based on the structure of the underlying specifications and the applied transformations. While the evaluation was not intended to produce statistical generalisations, it provides a robust qualitative foundation for assessing the method's reliability. Additional scenarios--such as the combination of noise with specific control-flow constructs--remain open for further exploration, but the current results already demonstrate the method's ability to produce interpretable and verifiable outcomes under varying conditions.



\section{Conclusion and future work}

This study presented an approach for automatically generating logical specifications from process models derived via workflow mining. Building on earlier work, the method integrates a pattern-based translation of process trees into temporal logic and leverages automated theorem proving for property validation. The extended experimental evaluation, which now includes both illustrative examples and real-case event logs from healthcare and financial domains, supports the practical feasibility of the approach and its applicability to realistic data.

The empirical results show that the workflow pattern base is expressive enough to model diverse process behaviours and that the generated logical specifications can be effectively tested for key properties. The proposed method supports multiple levels of validation, including satisfiability, logical relationships, and alignment with domain-specific requirements. Importantly, the approach proved robust when applied to real-world logs, enabling the identification and formal reasoning over relevant behavioural constraints.

Incorporating noise into the logs allowed for the exploration of data quality effects on model construction and logic generation. Although structural degradation was observed at higher noise levels, the method-particularly in its extended version-was still capable of capturing and validating key behavioural patterns. This suggests that the framework is not only suitable for idealised data but also viable in imperfect or incomplete real-world settings.

Future work will aim to increase the automation and scalability of the approach. One direction is to explore the use of native temporal logic solvers (e.g., PLTL) and dedicated proof strategies tailored to the structure of mined specifications. Another promising avenue involves learning requirement patterns directly from data or stakeholder input, thereby reducing the need for manual formalisation. Finally, applying the method to additional domains, including manufacturing and public services, would help further assess its generalisability and operational relevance.

\backmatter
%
%
%
%
%
\bmhead{Acknowledgements}

The authors gratefully acknowledge the technical support of
Jakub Grzyb and Gilbert Guszcza
during the experimental phase of this study.

%
%
%
%
\section*{Declarations}

\begin{itemize}
\item \textbf{Funding}
This research received no external funding.

\item \textbf{Conflict of interest}
The authors declare that they have no conflict of interest.

\item \textbf{Ethics approval and consent to participate}
Not applicable.

\item \textbf{Consent for publication}
Not applicable.

\item \textbf{Data availability}
The datasets used and analysed during the current study are publicly available from the sources referenced in the manuscript (see four footnotes and the link provided here: \url{https://home.agh.edu.pl/\~rklimek/doku.php?id=biglogs}).

\item \textbf{Materials availability}
Not applicable.

\item \textbf{Code availability}
The source code developed for this study is available from the corresponding author upon reasonable request.

\item \textbf{Author contribution}
Radoslaw Klimek: conceptualisation, methodology, formal analysis, supervision, writing-original draft.
Julia Witek: implementation of algorithms, contributions to the initial conference version of this work.
Authors approved the final manuscript.
\end{itemize}

\bibliography{../bib/bib-rk,../bib/bib-rk-main,../bib/bib-rk-tools}


\begin{thebibliography}{42}
\ifx \bisbn   \undefined \def \bisbn  #1{ISBN #1}\fi
\ifx \binits  \undefined \def \binits#1{#1}\fi
\ifx \bauthor  \undefined \def \bauthor#1{#1}\fi
\ifx \batitle  \undefined \def \batitle#1{#1}\fi
\ifx \bjtitle  \undefined \def \bjtitle#1{#1}\fi
\ifx \bvolume  \undefined \def \bvolume#1{\textbf{#1}}\fi
\ifx \byear  \undefined \def \byear#1{#1}\fi
\ifx \bissue  \undefined \def \bissue#1{#1}\fi
\ifx \bfpage  \undefined \def \bfpage#1{#1}\fi
\ifx \blpage  \undefined \def \blpage #1{#1}\fi
\ifx \burl  \undefined \def \burl#1{\textsf{#1}}\fi
\ifx \doiurl  \undefined \def \doiurl#1{\url{https://doi.org/#1}}\fi
\ifx \betal  \undefined \def \betal{\textit{et al.}}\fi
\ifx \binstitute  \undefined \def \binstitute#1{#1}\fi
\ifx \binstitutionaled  \undefined \def \binstitutionaled#1{#1}\fi
\ifx \bctitle  \undefined \def \bctitle#1{#1}\fi
\ifx \beditor  \undefined \def \beditor#1{#1}\fi
\ifx \bpublisher  \undefined \def \bpublisher#1{#1}\fi
\ifx \bbtitle  \undefined \def \bbtitle#1{#1}\fi
\ifx \bedition  \undefined \def \bedition#1{#1}\fi
\ifx \bseriesno  \undefined \def \bseriesno#1{#1}\fi
\ifx \blocation  \undefined \def \blocation#1{#1}\fi
\ifx \bsertitle  \undefined \def \bsertitle#1{#1}\fi
\ifx \bsnm \undefined \def \bsnm#1{#1}\fi
\ifx \bsuffix \undefined \def \bsuffix#1{#1}\fi
\ifx \bparticle \undefined \def \bparticle#1{#1}\fi
\ifx \barticle \undefined \def \barticle#1{#1}\fi
\bibcommenthead
\ifx \bconfdate \undefined \def \bconfdate #1{#1}\fi
\ifx \botherref \undefined \def \botherref #1{#1}\fi
\ifx \url \undefined \def \url#1{\textsf{#1}}\fi
\ifx \bchapter \undefined \def \bchapter#1{#1}\fi
\ifx \bbook \undefined \def \bbook#1{#1}\fi
\ifx \bcomment \undefined \def \bcomment#1{#1}\fi
\ifx \oauthor \undefined \def \oauthor#1{#1}\fi
\ifx \citeauthoryear \undefined \def \citeauthoryear#1{#1}\fi
\ifx \endbibitem  \undefined \def \endbibitem {}\fi
\ifx \bconflocation  \undefined \def \bconflocation#1{#1}\fi
\ifx \arxivurl  \undefined \def \arxivurl#1{\textsf{#1}}\fi
\csname PreBibitemsHook\endcsname

\bibitem[\protect\citeauthoryear{Mol and Primiero}{2015}]{DeMol-Primiero-2015}
\begin{barticle}
\bauthor{\bsnm{Mol}, \binits{L.D.}},
\bauthor{\bsnm{Primiero}, \binits{G.}}:
\batitle{When logic meets engineering: Introduction to logical issues in the
  history and philosophy of computer science}.
\bjtitle{History and Philosophy of Logic}
\bvolume{36}(\bissue{3}),
\bfpage{195}--\blpage{204}
(\byear{2015})
\doiurl{10.1080/01445340.2015.1084183}
\end{barticle}
\endbibitem

\bibitem[\protect\citeauthoryear{Broy}{2013}]{Broy-2013}
\begin{bchapter}
\bauthor{\bsnm{Broy}, \binits{M.}}:
\bctitle{On the role of logic and algebra in software engineering}.
In: \beditor{\bsnm{Paule}, \binits{P.}} (ed.)
\bbtitle{Mathematics, Computer Science and Logic -- A Never Ending Story: The
  Bruno Buchberger Festschrift},
pp. \bfpage{51}--\blpage{68}.
\bpublisher{Springer}, \blocation{???}
(\byear{2013}).
\doiurl{10.1007/978-3-319-00966-7\_2}
\end{bchapter}
\endbibitem

\bibitem[\protect\citeauthoryear{Klimek}{2019}]{Klimek-2019-LAMP}
\begin{barticle}
\bauthor{\bsnm{Klimek}, \binits{R.}}:
\batitle{Pattern-based and composition-driven automatic generation of logical
  specifications for workflow-oriented software models}.
\bjtitle{Journal of Logical and Algebraic Methods in Programming}
\bvolume{104},
\bfpage{201}--\blpage{226}
(\byear{2019})
\doiurl{10.1016/j.jlamp.2019.02.005}
\end{barticle}
\endbibitem

\bibitem[\protect\citeauthoryear{van~der Aalst}{2016}]{vanderAalst-2016}
\begin{bbook}
\bauthor{\bsnm{Aalst}, \binits{W.M.P.}}:
\bbtitle{Process Mining: Data Science in Action},
\bedition{2}nd edn.
\bpublisher{Springer}, \blocation{???}
(\byear{2016}).
\doiurl{10.1007/978-3-662-49851-4}
\end{bbook}
\endbibitem

\bibitem[\protect\citeauthoryear{Augusto et~al.}{2019}]{Augusto-etal-2019}
\begin{barticle}
\bauthor{\bsnm{Augusto}, \binits{A.}},
\bauthor{\bsnm{Conforti}, \binits{R.}},
\bauthor{\bsnm{Dumas}, \binits{M.}},
\bauthor{\bsnm{Rosa}, \binits{M.L.}},
\bauthor{\bsnm{Maggi}, \binits{F.M.}},
\bauthor{\bsnm{Marrella}, \binits{A.}},
\bauthor{\bsnm{Mecella}, \binits{M.}},
\bauthor{\bsnm{Soo}, \binits{A.}}:
\batitle{Automated discovery of process models from event logs: Review and
  benchmark}.
\bjtitle{IEEE Transactions on Knowledge and Data Engineering}
\bvolume{31}(\bissue{4}),
\bfpage{686}--\blpage{705}
(\byear{2019})
\doiurl{10.1109/TKDE.2018.2841877}
\end{barticle}
\endbibitem

\bibitem[\protect\citeauthoryear{Wi{\'s}niewski
  et~al.}{2019}]{Wisniewski-etal-2019}
\begin{bchapter}
\bauthor{\bsnm{Wi{\'s}niewski}, \binits{P.}},
\bauthor{\bsnm{Kluza}, \binits{K.}},
\bauthor{\bsnm{Jobczyk}, \binits{K.}},
\bauthor{\bsnm{Stachura-Terlecka}, \binits{B.}},
\bauthor{\bsnm{Ligeza}, \binits{A.}}:
\bctitle{Overview of generation methods for business process models}.
In: \bbtitle{Knowledge Science, Engineering and Management: 12th International
  Conference, KSEM 2019, Athens, Greece, August 28--30, 2019, Proceedings, Part
  II 12},
pp. \bfpage{55}--\blpage{60}
(\byear{2019}).
\bcomment{Springer}
\end{bchapter}
\endbibitem

\bibitem[\protect\citeauthoryear{van~der Aalst et~al.}{2004}]{vanderAalst-2004}
\begin{barticle}
\bauthor{\bsnm{Aalst}, \binits{W.}},
\bauthor{\bsnm{Weijters}, \binits{T.}},
\bauthor{\bsnm{Maruster}, \binits{L.}}:
\batitle{Workflow mining: discovering process models from event logs}.
\bjtitle{IEEE Transactions on Knowledge and Data Engineering}
\bvolume{16}(\bissue{9}),
\bfpage{1128}--\blpage{1142}
(\byear{2004})
\doiurl{10.1109/TKDE.2004.47}
\end{barticle}
\endbibitem

\bibitem[\protect\citeauthoryear{Weijters et~al.}{2006}]{Weijters-etal-2006}
\begin{bbook}
\bauthor{\bsnm{Weijters}, \binits{A.}},
\bauthor{\bsnm{Aalst}, \binits{W.}},
\bauthor{\bsnm{Medeiros}, \binits{A.}}:
\bbtitle{{Process Mining with the Heuristics Miner-algorithm}}.
\bpublisher{BETA Working Paper Series, WP 166},
\blocation{Eindhoven University of Technology, Eindhoven}
(\byear{2006})
\end{bbook}
\endbibitem

\bibitem[\protect\citeauthoryear{Leemans et~al.}{2013}]{Leemans-etal-2013}
\begin{bchapter}
\bauthor{\bsnm{Leemans}, \binits{S.J.J.}},
\bauthor{\bsnm{Fahland}, \binits{D.}},
\bauthor{\bsnm{Aalst}, \binits{W.M.P.}}:
\bctitle{Discovering block-structured process models from event logs containing
  infrequent behaviour}.
In: \beditor{\bsnm{Lohmann}, \binits{N.}},
\beditor{\bsnm{Song}, \binits{M.}},
\beditor{\bsnm{Wohed}, \binits{P.}} (eds.)
\bbtitle{Business Process Management Workshops - {BPM} 2013 International
  Workshops, Beijing, China, August 26, 2013, Revised Papers}.
\bsertitle{Lecture Notes in Business Information Processing},
vol. \bseriesno{171},
pp. \bfpage{66}--\blpage{78}.
\bpublisher{Springer}, \blocation{???}
(\byear{2013}).
\doiurl{10.1007/978-3-319-06257-0\_6}
\end{bchapter}
\endbibitem

\bibitem[\protect\citeauthoryear{Brzychczy et~al.}{2025}]{Brzychczy-etal-2024}
\begin{bchapter}
\bauthor{\bsnm{Brzychczy}, \binits{E.}},
\bauthor{\bsnm{Kluza}, \binits{K.}},
\bauthor{\bsnm{Sza{\l}a}, \binits{L.}}:
\bctitle{Enhancement of low-level event abstraction with large language models
  (llms)}.
In: \bbtitle{Business Process Management Workshops. BPM 2024 International
  Workshops, Krakow, Poland, September 1–6, 2024, Revised Selected Papers},
pp. \bfpage{209}--\blpage{220}
(\byear{2025}).
\bcomment{Springer}
\end{bchapter}
\endbibitem

\bibitem[\protect\citeauthoryear{Mendling et~al.}{2008}]{Mendling-etal-2008}
\begin{barticle}
\bauthor{\bsnm{Mendling}, \binits{J.}},
\bauthor{\bsnm{Lassen}, \binits{K.B.}},
\bauthor{\bsnm{Zdun}, \binits{U.}}:
\batitle{On the transformation of control flow between block-oriented and
  graph-oriented process modelling languages}.
\bjtitle{Int. J. Bus. Process. Integr. Manag.}
\bvolume{3}(\bissue{2}),
\bfpage{96}--\blpage{108}
(\byear{2008})
\doiurl{10.1504/IJBPIM.2008.020973}
\end{barticle}
\endbibitem

\bibitem[\protect\citeauthoryear{Ouyang et~al.}{2006}]{Ouyang-etal-2006b}
\begin{bchapter}
\bauthor{\bsnm{Ouyang}, \binits{C.}},
\bauthor{\bsnm{Dumas}, \binits{M.}},
\bauthor{\bsnm{Hofstede}, \binits{A.H.M.}},
\bauthor{\bsnm{Aalst}, \binits{W.M.P.}}:
\bctitle{From bpmn process models to bpel web services}.
In: \bbtitle{IEEE International Conference on Web Services (ICWS'06)},
pp. \bfpage{285}--\blpage{292}
(\byear{2006})
\end{bchapter}
\endbibitem

\bibitem[\protect\citeauthoryear{Recker and
  Mendling}{2006}]{Recker-Mendling-2006}
\begin{bchapter}
\bauthor{\bsnm{Recker}, \binits{J.C.}},
\bauthor{\bsnm{Mendling}, \binits{J.}}:
\bctitle{On the translation between bpmn and bpel: Conceptual mismatch between
  process modeling languages}.
In: \beditor{\bsnm{Latour}, \binits{T.}},
\beditor{\bsnm{Petit}, \binits{M.}} (eds.)
\bbtitle{18th International Conference on Advanced Information Systems
  Engineering},
pp. \bfpage{521}--\blpage{532}.
\bpublisher{Namur University Press}, \blocation{???}
(\byear{2006})
\end{bchapter}
\endbibitem

\bibitem[\protect\citeauthoryear{{van der Aalst} and {Bisgaard
  Lassen}}{2008}]{vanderAalst-BisgaardLassen-2008}
\begin{barticle}
\bauthor{\bsnm{{van der Aalst}}, \binits{W.M.P.}},
\bauthor{\bsnm{{Bisgaard Lassen}}, \binits{K.}}:
\batitle{Translating unstructured workflow processes to readable bpel: Theory
  and implementation}.
\bjtitle{Information and Software Technology}
\bvolume{50}(\bissue{3}),
\bfpage{131}--\blpage{159}
(\byear{2008})
\doiurl{10.1016/j.infsof.2006.11.004}
\end{barticle}
\endbibitem

\bibitem[\protect\citeauthoryear{van Zelst and
  Leemans}{2020}]{vanZelst-Leemans-2020}
\begin{botherref}
\oauthor{\bsnm{Zelst}, \binits{S.J.}},
\oauthor{\bsnm{Leemans}, \binits{S.J.J.}}:
Translating workflow nets to process trees: An algorithmic approach.
Algorithms
\textbf{13}(11)
(2020)
\doiurl{10.3390/a13110279}
\end{botherref}
\endbibitem

\bibitem[\protect\citeauthoryear{Ferilli}{2016}]{Ferilli-2016}
\begin{bchapter}
\bauthor{\bsnm{Ferilli}, \binits{S.}}:
\bctitle{The woman formalism for expressing process models}.
In: \beditor{\bsnm{Perner}, \binits{P.}} (ed.)
\bbtitle{Advances in Data Mining. Applications and Theoretical Aspects},
pp. \bfpage{363}--\blpage{378}.
\bpublisher{Springer},
\blocation{Cham}
(\byear{2016})
\end{bchapter}
\endbibitem

\bibitem[\protect\citeauthoryear{Roubtsova}{2005}]{Roubtsova-2005}
\begin{bchapter}
\bauthor{\bsnm{Roubtsova}, \binits{E.E.}}:
\bctitle{Property driven mining in workflow logs}.
In: \beditor{\bsnm{K{\l}opotek}, \binits{M.A.}},
\beditor{\bsnm{Wierzcho{\'{n}}}, \binits{S.T.}},
\beditor{\bsnm{Trojanowski}, \binits{K.}} (eds.)
\bbtitle{Intelligent Information Processing and Web Mining},
pp. \bfpage{471}--\blpage{475}.
\bpublisher{Springer},
\blocation{Berlin, Heidelberg}
(\byear{2005})
\end{bchapter}
\endbibitem

\bibitem[\protect\citeauthoryear{Emerson}{1990}]{Emerson-1990}
\begin{bchapter}
\bauthor{\bsnm{Emerson}, \binits{E.A.}}:
\bctitle{Temporal and modal logic}.
In: \beditor{\bsnm{Leeuwen}, \binits{J.}} (ed.)
\bbtitle{Handbook of Theoretical Computer Science}
vol. \bseriesno{B},
pp. \bfpage{995}--\blpage{1072}.
\bpublisher{MIT Press}, \blocation{???}
(\byear{1990}).
\burl{http://dl.acm.org/citation.cfm?id=114891.114907}
\end{bchapter}
\endbibitem

\bibitem[\protect\citeauthoryear{Manna and Pnueli}{1992}]{Manna-Pnueli-1992}
\begin{bbook}
\bauthor{\bsnm{Manna}, \binits{Z.}},
\bauthor{\bsnm{Pnueli}, \binits{A.}}:
\bbtitle{The Temporal Logic of Reactive and Concurrent Systems --
  Specification}.
\bpublisher{Springer}, \blocation{???}
(\byear{1992})
\end{bbook}
\endbibitem

\bibitem[\protect\citeauthoryear{van~der Aalst
  et~al.}{2005}]{vanderAalst-etal-2005}
\begin{bchapter}
\bauthor{\bsnm{Aalst}, \binits{W.M.P.}},
\bauthor{\bsnm{Beer}, \binits{H.T.}},
\bauthor{\bsnm{Dongen}, \binits{B.F.}}:
\bctitle{Process mining and verification of properties: An approach based on
  temporal logic}.
In: \beditor{\bsnm{Meersman}, \binits{R.}},
\beditor{\bsnm{Tari}, \binits{Z.}} (eds.)
\bbtitle{On the Move to Meaningful Internet Systems 2005: CoopIS, DOA, and
  ODBASE},
pp. \bfpage{130}--\blpage{147}.
\bpublisher{Springer},
\blocation{Berlin, Heidelberg}
(\byear{2005})
\end{bchapter}
\endbibitem

\bibitem[\protect\citeauthoryear{Klimek and
  Witek}{2024}]{Klimek-Witek-2024-ASE-RENE}
\begin{bchapter}
\bauthor{\bsnm{Klimek}, \binits{R.}},
\bauthor{\bsnm{Witek}, \binits{J.}}:
\bctitle{Automatic generation of logical specifications for behavioural
  models}.
In: \bbtitle{Proceedings of the 39th IEEE/ACM International Conference on
  Automated Software Engineering Workshops (ASE/RENE), Sun 27 October--Fri 1
  November 2024, Sacramento, CA, USA}.
\bsertitle{ASEW'24},
pp. \bfpage{1}--\blpage{7}.
\bpublisher{Association for Computing Machinery},
\blocation{New York, NY, USA}
(\byear{2024}).
\doiurl{10.1145/3695750.3695822} .
\burl{https://doi.org/10.1145/3695750.3695822}
\end{bchapter}
\endbibitem

\bibitem[\protect\citeauthoryear{Szab{\'o}}{2012}]{Szabo-2012}
\begin{bchapter}
\bauthor{\bsnm{Szab{\'o}}, \binits{Z.G.}}:
\bctitle{Compositionality}.
In: \beditor{\bsnm{Zalta}, \binits{E.N.}} (ed.)
\bbtitle{Stanford Encyclopedia of Philosophy
  {\texttt{HTTP://PLATO.STANFORD.EDU/ENTRIES/COMPOSITIONALITY/}}}
(\byear{2012}).
\bcomment{accessed on 20-Feb-2018}
\end{bchapter}
\endbibitem

\bibitem[\protect\citeauthoryear{Recanati}{2012}]{Recanti-2012}
\begin{bchapter}
\bauthor{\bsnm{Recanati}, \binits{F.}}:
\bctitle{Composition ality, flexibility, and context dependence}.
In: \beditor{\bsnm{Werning}, \binits{M.}},
\beditor{\bsnm{Hinzen}, \binits{W.}},
\beditor{\bsnm{Machery}, \binits{E.}} (eds.)
\bbtitle{The Oxford Handbook of Compositionality},
pp. \bfpage{175}--\blpage{191}.
\bpublisher{Oxford Handbook Online}, \blocation{???}
(\byear{2012}).
\doiurl{10.1093/oxfordhb/9780199541072.013.0008}
\end{bchapter}
\endbibitem

\bibitem[\protect\citeauthoryear{Tripakis}{2016}]{Tripakis-2016}
\begin{barticle}
\bauthor{\bsnm{Tripakis}, \binits{S.}}:
\batitle{Compositionality in the science of system design}.
\bjtitle{Proceedings of the {IEEE}}
\bvolume{104}(\bissue{5}),
\bfpage{960}--\blpage{972}
(\byear{2016})
\doiurl{10.1109/JPROC.2015.2510366}
\end{barticle}
\endbibitem

\bibitem[\protect\citeauthoryear{Bogar{\'i}n et~al.}{2018}]{Bogarn-etal-2018}
\begin{barticle}
\bauthor{\bsnm{Bogar{\'i}n}, \binits{A.}},
\bauthor{\bsnm{Cerezo}, \binits{R.}},
\bauthor{\bsnm{Romero}, \binits{C.}}:
\batitle{Discovering learning processes using inductive miner: A case study
  with learning management systems (lmss)}.
\bjtitle{Psicothema}
\bvolume{30},
\bfpage{322}--\blpage{329}
(\byear{2018})
\end{barticle}
\endbibitem

\bibitem[\protect\citeauthoryear{Wen et~al.}{2010}]{Wen-etal-2010}
\begin{barticle}
\bauthor{\bsnm{Wen}, \binits{L.}},
\bauthor{\bsnm{Wang}, \binits{J.}},
\bauthor{\bsnm{Aalst}, \binits{W.M.P.}},
\bauthor{\bsnm{Huang}, \binits{B.}},
\bauthor{\bsnm{Sun}, \binits{J.}}:
\batitle{Mining process models with prime invisible tasks}.
\bjtitle{Data \& Knowledge Engineering}
\bvolume{69}(\bissue{10}),
\bfpage{999}--\blpage{1021}
(\byear{2010})
\doiurl{10.1016/j.datak.2010.06.001}
\end{barticle}
\endbibitem

\bibitem[\protect\citeauthoryear{Leemans et~al.}{2024}]{Leemans-etal-2024}
\begin{barticle}
\bauthor{\bsnm{Leemans}, \binits{S.J.J.}},
\bauthor{\bsnm{Maggi}, \binits{F.M.}},
\bauthor{\bsnm{Montali}, \binits{M.}}:
\batitle{Enjoy the silence: Analysis of stochastic petri nets with silent
  transitions}.
\bjtitle{Information Systems}
\bvolume{124},
\bfpage{102383}
(\byear{2024})
\doiurl{10.1016/J.IS.2024.102383}
\end{barticle}
\endbibitem

\bibitem[\protect\citeauthoryear{Klimek}{2014}]{Klimek-2014-AMCS}
\begin{barticle}
\bauthor{\bsnm{Klimek}, \binits{R.}}:
\batitle{A system for deduction-based formal verification of workflow-oriented
  software models}.
\bjtitle{International Journal of Applied Mathematics and Computer Science}
\bvolume{24}(\bissue{4}),
\bfpage{941}--\blpage{956}
(\byear{2014})
\doiurl{10.2478/amcs-2014-0069}
\end{barticle}
\endbibitem

\bibitem[\protect\citeauthoryear{Wolter and
  Wooldridge}{2011}]{Wolter-Wooldridge-2011}
\begin{barticle}
\bauthor{\bsnm{Wolter}, \binits{F.}},
\bauthor{\bsnm{Wooldridge}, \binits{M.}}:
\batitle{Temporal and dynamic logic}.
\bjtitle{Journal of Indian Council of Philosophical Research}
\bvolume{XXVII(1)},
\bfpage{249}--\blpage{276}
(\byear{2011})
\end{barticle}
\endbibitem

\bibitem[\protect\citeauthoryear{Kleene}{1952}]{Kleene-1952}
\begin{bbook}
\bauthor{\bsnm{Kleene}, \binits{S.C.}}:
\bbtitle{Introduction to Metamathematics}.
\bsertitle{Bibliotheca Mathematica}.
\bpublisher{North-Holland}, \blocation{???}
(\byear{1952})
\end{bbook}
\endbibitem

\bibitem[\protect\citeauthoryear{van Benthem}{1993--95}]{vanBenthem-1995}
\begin{bchapter}
\bauthor{\bsnm{Benthem}, \binits{J.}}:
\bctitle{Temporal Logic}.
\bbtitle{Handbook of Logic in Artificial Intelligence and Logic Programming}.
\bsertitle{4},
pp. \bfpage{241}--\blpage{350}.
\bpublisher{Clarendon Press}, \blocation{???}
(\byear{1993--95})
\end{bchapter}
\endbibitem

\bibitem[\protect\citeauthoryear{Schulz}{2020}]{E-tool}
\begin{botherref}
\oauthor{\bsnm{Schulz}, \binits{S.}}:
Website for prover {E}.
accessed on 5-Aug-2024
(2020).
\url{http://wwwlehre.dhbw-stuttgart.de/\textasciitilde sschulz/E/E.html}
\end{botherref}
\endbibitem

\bibitem[\protect\citeauthoryear{Schulz}{2002}]{Schulz-2002}
\begin{barticle}
\bauthor{\bsnm{Schulz}, \binits{S.}}:
\batitle{E -- a brainiac theorem prover}.
\bjtitle{Journal of AI Communications}
\bvolume{15}(\bissue{2,3}),
\bfpage{111}--\blpage{126}
(\byear{2002})
\end{barticle}
\endbibitem

\bibitem[\protect\citeauthoryear{Voronkov}{2017}]{Vampire-tool}
\begin{botherref}
\oauthor{\bsnm{Voronkov}, \binits{A.}}:
Website for prover {Vampire}.
accessed on 5-Aug-2024
(2017).
\url{https://vprover.github.io/}
\end{botherref}
\endbibitem

\bibitem[\protect\citeauthoryear{Riazanov and
  Voronkov}{2002}]{Riazanov-Voronkov-2002}
\begin{barticle}
\bauthor{\bsnm{Riazanov}, \binits{A.}},
\bauthor{\bsnm{Voronkov}, \binits{A.}}:
\batitle{The design and implementation of {VAMPIRE}}.
\bjtitle{Journal of AI Communications}
\bvolume{15}(\bissue{2,3}),
\bfpage{91}--\blpage{110}
(\byear{2002})
\end{barticle}
\endbibitem

\bibitem[\protect\citeauthoryear{Sutcliffe}{2017}]{Sutcliffe-2017}
\begin{barticle}
\bauthor{\bsnm{Sutcliffe}, \binits{G.}}:
\batitle{{The TPTP Problem Library and Associated Infrastructure}}.
\bjtitle{Journal of Automated Reasoning}
\bvolume{59},
\bfpage{438}--\blpage{502}
(\byear{2017})
\end{barticle}
\endbibitem

\bibitem[\protect\citeauthoryear{Schreiner}{2023}]{Schreiner-2023}
\begin{botherref}
\oauthor{\bsnm{Schreiner}, \binits{W.}}:
First-order logic: software for proving, Course 'Computational logic',
  \url{https://moodle.risc.jku.at/pluginfile.php/11902/mod_resource/content/9/10-fol6.pdf}.
  The last access 25.04.2024
(2023)
\end{botherref}
\endbibitem

\bibitem[\protect\citeauthoryear{Lamport}{1977}]{Lamport-1977}
\begin{barticle}
\bauthor{\bsnm{Lamport}, \binits{L.}}:
\batitle{Proving the correctness of multiprocess programs}.
\bjtitle{IEEE Transactions on Software Engineering}
\bvolume{3}(\bissue{2}),
\bfpage{125}--\blpage{143}
(\byear{1977})
\doiurl{10.1109/TSE.1977.229904}
\end{barticle}
\endbibitem

\bibitem[\protect\citeauthoryear{Alpern and
  Schneider}{1985}]{Alpern-Schneider-1985}
\begin{barticle}
\bauthor{\bsnm{Alpern}, \binits{B.}},
\bauthor{\bsnm{Schneider}, \binits{F.B.}}:
\batitle{Defining liveness}.
\bjtitle{Information Processing Letters}
\bvolume{21 (4)},
\bfpage{181}--\blpage{185}
(\byear{1985})
\end{barticle}
\endbibitem

\bibitem[\protect\citeauthoryear{Kindler}{1994}]{Kindler-1994}
\begin{botherref}
\oauthor{\bsnm{Kindler}, \binits{E.}}:
Safety and liveness properties: A survey.
EATCS-Bulletin
\textbf{53}
(1994)
\end{botherref}
\endbibitem

\bibitem[\protect\citeauthoryear{Ben-Ari}{2012}]{Ben-Ari-2012}
\begin{bbook}
\bauthor{\bsnm{Ben-Ari}, \binits{M.}}:
\bbtitle{Mathematical Logic for Computer Science},
\bedition{3rd} edn.
\bpublisher{Springer}, \blocation{???}
(\byear{2012})
\end{bbook}
\endbibitem

\bibitem[\protect\citeauthoryear{Klimek and
  Semczyszyn}{2025}]{Klimek-Semczyszyn-2025-EASE}
\begin{bchapter}
\bauthor{\bsnm{Klimek}, \binits{R.}},
\bauthor{\bsnm{Semczyszyn}}:
\bctitle{Re-evaluation of logical specification in behavioural verification}.
In: \bbtitle{Proceedings of the 29th International Conference on Evaluation and
  Assessment in Software Engineering (EASE 2025), 17--20 June, 2025, Istanbul,
  Turkey. Search Also Preprints, \url{http://arxiv.org/abs/2505.17979}}
(\byear{2025})
\end{bchapter}
\endbibitem

\end{thebibliography}

\end{document}